\documentclass{article}
\usepackage[utf8]{inputenc}
\usepackage[T1]{fontenc}
\usepackage[english]{babel} 
\PassOptionsToPackage{hyphens}{url}
\usepackage{babel}
\usepackage{amsmath}
\usepackage{graphicx}
\usepackage{fancyhdr}
\newcommand{\scidatalogo}{\includegraphics[height=36pt]{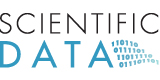}}
\newcommand{\overleaflogo}{\includegraphics[height=36pt]{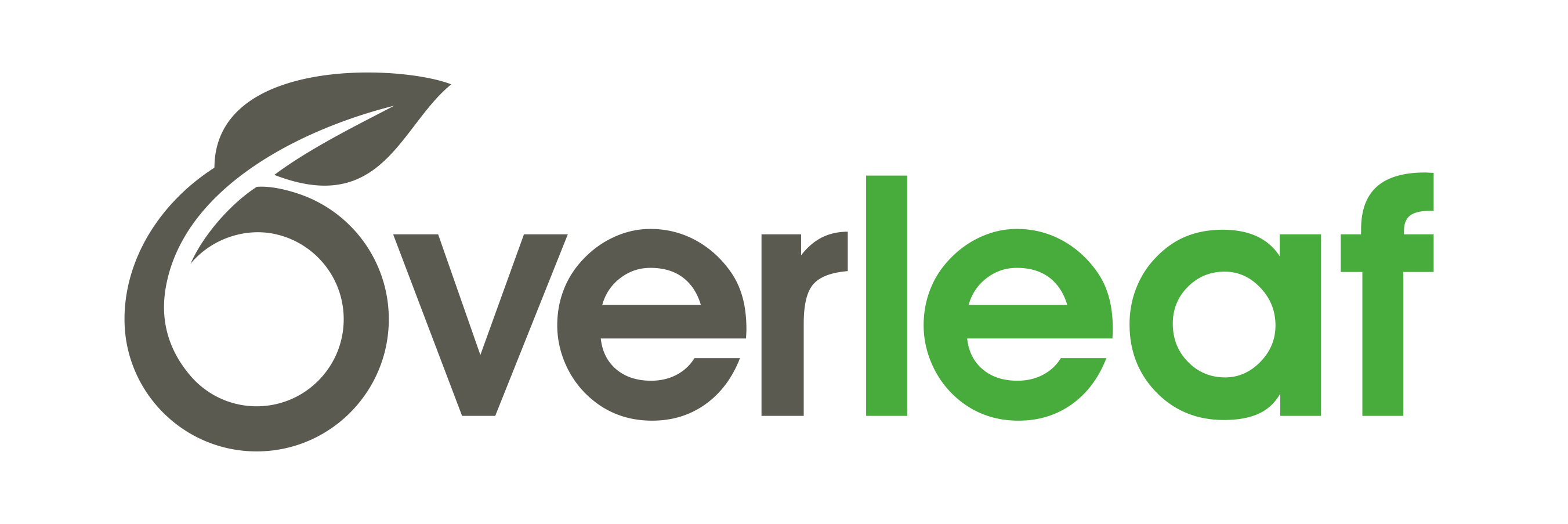}}
\usepackage{color,soul}
\usepackage{siunitx}
\usepackage{tabularx}
\usepackage{tabulary}
\usepackage{booktabs}
\usepackage{array}

\newcolumntype{C}[1]{>{\centering\arraybackslash\hspace{0pt}}p{#1}}
\usepackage{graphicx}
\usepackage{subcaption}
\usepackage{pdflscape}
\usepackage[colorlinks=true]{hyperref}
\usepackage{xcolor}
\hypersetup{colorlinks=true,linkcolor = blue,urlcolor  = blue,
            citecolor = blue,
             anchorcolor = blue}

\usepackage{adjustbox}
\usepackage{longtable}
\usepackage{float}
\usepackage{rotating}
\usepackage{lineno}

\begin{document}

\title{Harmonised LUCAS in-situ data and photos on land cover and use from 5 tri-annual surveys in the European Union}
%\title{15-years of collecting in-situ data on land cover and use - harmonizing the LUCAS survey in the European Union}

\author{Raphaël d'Andrimont\textsuperscript{1{*}},
Momchil Yordanov\textsuperscript{1},\\
Laura Martinez-Sanchez\textsuperscript{1},
Beatrice Eiselt\textsuperscript{2},
Alessandra Palmieri\textsuperscript{2},\\
Paolo Dominici\textsuperscript{2},
Javier Gallego\textsuperscript{1},
Hannes Isaak Reuter\textsuperscript{2},\\
Christian Joebges\textsuperscript{3}, 
Guido Lemoine\textsuperscript{1},
Marijn van der Velde\textsuperscript{1{*}}
}

% emails
% raphael.dandrimont@ec.europa.eu
% Momtchil.IORDANOV@ext.ec.europa.eu
% Guido.LEMOINE@ec.europa.eu
% Beatrice.EISELT@ec.europa.eu
% Alessandra.Palmieri@ec.europa.eu
% paolo.dominici@ec.europa.eu
% jagalpi@gmail.com
% Hannes.reuter@ec.europa.eu
% Christian.Joebges@gopa.lu
% Laura.Martinez-Sanchez@ec.europa.eu
% marijn.van-der-velde@ec.europa.eu

\maketitle
\thispagestyle{fancy}

1. European Commission Joint Research Centre (JRC), Ispra, Italy 

2. European Commission, Eurostat (ESTAT), Luxembourg, Luxembourg

3. GOPA, Luxembourg, Luxembourg

% 3. European Commission, Joint Research Centre (JRC) {---} Knowledge for Sustainable Development & Food Security Unit, Ispra , Italy  %4. ESA 

{*}corresponding author(s):\\
    Raphaël d'Andrimont (raphael.dandrimont@ec.europa.eu) and Marijn van der Velde (marijn.van-der-velde@ec.europa.eu)

% KEYWORDS
% statistics, agriculture, big data, land cover, land use, Copernicus, remote sensing, water management, soil, in-situ, cropland, eurostat, estat, ground truth

\begin{abstract}
% This is a manuscript template for Data Descriptor submissions to \emph{Scientific
% Data} (http://www.nature.com/scientificdata). The abstract must be
% no longer than 170 words, and should succinctly describe the study,
% the assay(s) performed, the resulting data, and the reuse potential,
% but should not make any claims regarding new scientific findings.
% No references are allowed in this section.
 
Accurately characterizing land surface changes with Earth Observation requires geo-localized ground truth. In the European Union (EU), a tri-annual surveyed sample of land cover and land use has been collected since 2006 under the Land Use/Cover Area frame Survey (LUCAS). A total of \num{1351293} observations at \num{651780} unique locations for 117 variables along with \num{5.4} million photos were collected during five LUCAS surveys. Until now, these data have never been harmonised into one database, limiting full exploitation of the information. This paper describes the LUCAS point sampling/surveying methodology, including collection of standard variables such as land cover, environmental parameters, and full resolution landscape and point photos, and then describes the harmonisation process. The resulting harmonised database is the most comprehensive in-situ dataset on land cover and use in the EU. The database is valuable for geo-spatial and statistical analysis of land use and land cover change. Furthermore, its potential to provide multi-temporal in-situ data will be enhanced by recent computational advances such as deep learning.  %, training and validation for Earth Observation, and image analysis and computer vision.  

\end{abstract}

\section{Background \& Summary}\label{sec:backgroundNsummary}

% (700 words maximum) An overview of the study design, the assay(s)
% performed, and the created data, including any background information
% needed to put this study in the context of previous work and the literature.
% The section should also briefly outline the broader goals that motivated
% the creation of this dataset and the potential reuse value. We also
% encourage authors to include a figure that provides a schematic overview
% of the study and assay(s) design. This section and the other main
% body sections of the manuscript should include citations to the literature
% as needed \cite{cite1, cite2}. References should be included within the 
% manuscript file itself as our system cannot accept BibTeX bibliography files. 
% Authors who wish to use BibTeX to prepare their references should therefore 
% copy the reference list from the .bbl file that BibTeX generates and paste it 
% into the main manuscript .tex file (and delete the associated 
% \textbackslash{}bibliography and \textbackslash{}bibliographystyle commands).

Accurate, timely, and representative in-situ observations across large areas have always been needed to report statistics on land use, land cover, and the environment. The Land Use/Cover Area frame Survey (LUCAS) in the European Union (EU) was set-up for this purpose \cite{gallego2010european,LUCAS:online}. Precise geo-localized in-situ information is also indispensable to train and validate algorithms that characterize the Earth's surface based on remotely sensed observations. Comprehensive and thematically rich in-situ data can lead to better classifiers and more accurate multi-temporal land surface mapping. This is especially true since increasingly frequent and detailed Earth Observations are being made, for instance by the fleet of Sentinel satellites of the EU's Copernicus space program. These developments are opening avenues to better combine classical statistical surveying and Earth Observation (EO) derived products in the domains of land use and land cover change and environmental monitoring (e.g. \cite{johnson2019using}). 

%More recently, in-situ observations are getting traction as the complement satellite data to address these issues. 
%Additionally, recent advances in computer vision provides a unique opportunity to extract information from historical photos taken in-situ.

%Classical field surveys to collect such in-situ data are costly. 
The Land Use/Cover Area frame Survey (LUCAS) is a 3-yearly in-situ land cover and land use data collection exercise that extends over the whole of the EU’s territory \cite{gallego2010european,LUCAS:online}. Here we present a harmonised, consolidated, user-friendly database to structure the wealth of observations and data collected during five LUCAS surveys with slightly different collection protocols in 2006, 2009, 2012, 2015 and 2018. 
LUCAS collects information on land cover and land use, agro-environmental variables, soil, and grassland. The surveys also provide territorial information to analyse the interactions between agriculture, environment, and countryside, such as irrigation and land management. LUCAS is based on a systematic sampling plan that allows for unbiased statistical land cover and land use area estimators throughout the EU territory. It uses a well documented harmonized classification with separate land cover and land use codes. Data quality is assured by a regular two-level quality control, in which all points are evaluated by quality controllers. %The survey was thus initially not designed for mapping. %Consistent and continuous in-situ observations are needed to calibrate and improve relevant crop information, but also to verify maps and support monitoring activities in near-real time.  

The LUCAS project was implemented following Decision 1445/2000/EC of the European Parliament and of the Council of 22 May 2000 “On the application of area-frame survey and remote-sensing techniques to the agricultural statistics for 1999 to 2003” and has continued since. %It continued until 2007 by Decision 2066/2003/EC of 10 November 2003, extended to EU-N10 by Decision 786/2004/EC of 21 April 2004. 
While the LUCAS survey concept was initiated and tested in 2001 and 2003 \cite{bettio2002area}, it has been restructured in 2006 \cite{gallego2005stratified} and then slightly modified to result in the actual survey design \cite{scarno2018lucas}.
%Since 2006, with a pilot on 11 countries, EUROSTAT has carried out LUCAS surveys every three years with the survey design ever evolving, however the core components have remained comparable. From the LUCAS survey  data collection, different types of information are available:
In 2006, Eurostat, the statistical office of the EU, launched a pilot survey project in 11 countries to test the stratified sampling design. The primary focus was on agricultural areas with more emphasis given to easily accessible points. Since then Eurostat has carried out LUCAS surveys every three years with the survey design ever evolving, however the LUCAS \textit{core} component (i.e.  the identification of the point, and the surveying of specific variables on different aspects of land cover, land use, and land and water management \cite{c12018}), has remained comparable for all five surveys. At each LUCAS point, standard variables are collected including land cover, land use, environmental parameters, as well as one downward facing photo of the point (P) and four landscape photos in the cardinal compass directions (N, E, S, W). From the LUCAS survey data collection, different types of information are available:

\begin{itemize}
\item The \textit{micro data} provides land cover, land use, and environmental variables associated to each surveyed point. 
\item The full-resolution point and landscape \textit{photos} taken in the four cardinal directions at each point.
\end{itemize}

% In addition to information collected, statistical tables with aggregated results are computed by land cover and land use at geographical level. %Extraced from the Ballin Paper

Additionally to the \textit{core} variables collected, other specific \textit{modules} were carried out on demand such as (i) the transect of 250 m to assess transitions of land cover and existing linear features (2009, 2012, 2015), (ii) the topsoil module (2009, 2012 (partly), 2015 and 2018), (iii) the grassland (2018), and (iv) the Copernicus module collecting the homogeneous extent of land cover on a 50-m radius (2018). Due to the specificity of these modules, their corresponding collected data are not included in the data harmonisation presented in this paper. The topsoil module datasets for 2009, 2012 and 2015 were harmonised resulting in the largest harmonised open‐access dataset of topsoil properties in the EU  \cite{orgiazzi2018lucas}.

LUCAS is a two phase sample survey. The LUCAS first phase sample is a systematic selection of points on a grid  with a 2 km spacing in Eastings and Northings covering the whole of the EU’s territory \cite{gallego2008using}. Currently, it includes around 1.1 million points (Figure \ref{fig:harmo}),and is referred to as the master sample. Each point of the first phase sample is classified in one of ten land‐cover classes via photo interpretation of ortho-photos or satellite images \cite{LUCAS-StratificationGuidelines:online}. 

While the core sampling and survey methods have remained the same throughout the five surveys, evolving goals of the surveys have lead to slightly different sample allocations for different land covers. In 2006, the main objective was to 'make early estimates of the main crop areas', along with the ability to collect information on agri-environmental indicators in the context of the monitoring of the Common Agriculture Policy (CAP) \cite{bettio2002area}. In  2009, the main  objective was to estimate areas, especially in conjunction with other data sources such as Corine Land Cover (CLC) \cite{gallego2008using}. In 2018, the main objective was to 'monitor social and economic use of land as well as ecosystems and biodiversity' \cite{scarno2018lucas}. Additionally, in 2018, a linear logistic regression model based on LUCAS 2015 and 16 additional variables where used as co-variates to forecast the most probable land cover for each of these points \cite{scarno2018lucas}. From this stratified first phase sample, the second phase sample of points is selected to obtain the desired statistically representative spatial distribution of sampled land cover classes. %In 2018, the second phase stratification is not fixed, but rather obtained in a dynamic way \cite{scarno2018lucas}. 
% Essentially, the 'atomic strata' (Cartesian product of STR18 (grid classification), CLC, and elevation classifications) are being aggregated by an iterative algorithm that optimises the coefficients of variance of the target variables, whilst also taking into account the related, desired sampling errors. 
With LUCAS 2018 this amounts to \num{337845} points, out of which approximately \num{240000} points are visited in the field by surveyors to collect additional information that cannot be assessed remotely. 

The in-situ nature of the survey implies that the majority of the data (73\% over the five surveys, for details see Table \ref{tab:pointsPerCountry} below) are gathered through direct observations made by surveyors on the ground. However, points unlikely to change and points too difficult to access are assigned by photo-interpretation in the office (16\% over the five surveys), using the latest available ortho-photos or Very High Resolution (VHR) images.
Although most of the points assigned for in-situ assessment can be visited in the field, about 11\% of the points cannot be reached (e.g. no access, point is at more than 30 minutes walking distance) and are photo-interpreted on ortho-photos or Very High Resolution (VHR) images in the field. Furthermore, while the observations are associated to the location of the theoretical LUCAS point, the actual geo-location of the point from which the photos were taken needs to be derived from the GPS information collected by the surveyor. 
In the scientific literature, LUCAS land cover and land use survey data have been used to derive statistical estimates \cite{gallego2010european}, to describe land cover/use diversity at regional level \cite{palmieri2011land}, and its sampling frame was used as a basis for various applications including assessing the availability of crowd-sourced photos potentially relevant for crop monitoring across the EU \cite{d2018crowdsourced}. LUCAS was designed to derive statistics for area estimation (e.g. \cite{bettio2002area} and \cite{gallego2008using}). 
Recently, several researchers have started to use LUCAS data in large scale land cover mapping processes, especially as a source of training and/or validation data for supervised classification approaches at regional/national scale \cite{karydas2015use,mack2017semi,close2018use,pflugmacher2019mapping,leinenkugel2019potential,d2020detecting,weigand2020spatial}. 

% The need for harmonisation 
Several drawbacks become apparent when working with the original LUCAS datasets. While the inconsistencies could be due to the enumerators’ subjectivity in interpretation of the legends and the legend itself, it is also related to the complexity of the field survey:  large number of surveyors ($>$ 700), complex documentation for the enumerators ($>$ 400 pages combining all the documents), translated to 20 languages. These drawbacks hinder the further use of the LUCAS data by the scientific community as a whole and in particular those active in emerging fields of big data analytics, data fusion, and computer vision. They include:
\begin{itemize}
\item Inconsistencies and errors between legends and labels from one LUCAS survey to the next which is hampering temporal analysis.
\item Missing internal cross-references in the datasets that would facilitate computation and linking observed variables, photos, etc.
\item The original full resolution photos taken at each surveyed point are not available for download. 
\item The lack of a single-entry point or consolidated database hampering automated processing and big data analysis.

\end{itemize}
 
Therefore, we have gone through an extensive process of cleaning, semantic and topological harmonisation, along with connecting the originally disjoint LUCAS datasets in one consolidated database with hard-coded links to the full-resolution photos, openly accessible along with this paper.

%Yet, there is a growing need for consistent and continuous in-situ data to calibrate and improve spatial information, but also to verify maps and support monitoring activities in near-real time. So far,

\section{Methods}

% The Methods should include detailed text describing any steps or procedures used in producing the data, including full descriptions of the experimental design, data acquisition assays, and any computational processing (e.g. normalization, image feature extraction). See the detailed section in our submission guidelines for advice on writing a transparent and reproducible methods section. Related methods should be grouped under corresponding subheadings where possible, and methods should be described in enough detail to allow other researchers to interpret and repeat, if required, the full study. Specific data outputs should be explicitly referenced via data citation (see Data Records and Citing Data, below).
% Authors should cite previous descriptions of the methods under use, but ideally the method descriptions should be complete enough for others to understand and reproduce the methods and processing steps without referring to associated publications. There is no limit to the length of the Methods section.

\begin{figure}[!h]
    \centering
    %\hspace*{-2.5cm}     
    \includegraphics[width=1\textwidth]{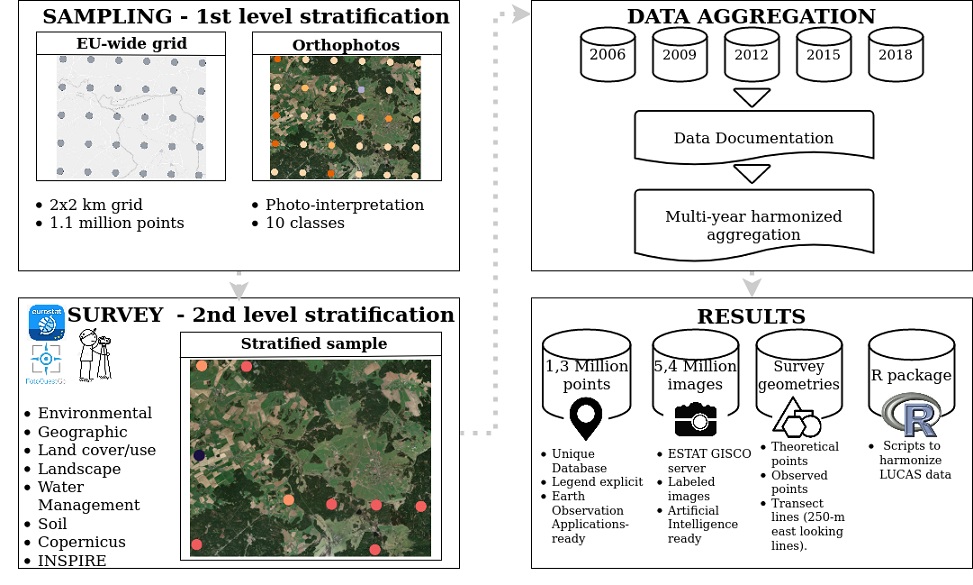}
    \caption{Schematic overview of the LUCAS and harmonisation methodologies. The left side illustrates the sampling at the basis of the production of the LUCAS primary data. The top right side shows the raw base data (micro data). The process of harmonising is contained within the multi-year harmonised aggregation block and is the subject of the following two sections. The bottom right presents the two outputs of this study (more in section \ref{sec:dataRecords}) - a harmonised, legend-explicit, multi-year, user-friendly, version of the LUCAS micro data (section \ref{sec:o_survey}), and a database with all cardinal-direction landscape and point photos collected during the surveys, including their respective EXIF attributes (section \ref{sec:o_pictures}).}
    \label{fig:harmo}
\end{figure}

Having contextualized the LUCAS survey, we proceed with describing the full methodological workflow to harmonize the data, as schematically shown in Figure \ref{fig:harmo}. The Sampling and Survey sub-figures provide an overview of the methodological framework of the LUCAS data collection itself (see previous section \ref{sec:backgroundNsummary}). The Data aggregation and Results sub-figures illustrate the work carried out in this study. The datasets collected during the five surveys (in 2006, 2009, 2012, 2015, 2018) are the main LUCAS products available (more in section \ref{sec:protocol1}). These datasets and their respective data documentation were used to create the multi-year harmonised database. The harmonisation process is described below and in Table \ref{tab:methodMicroData}. The results are consolidated in one unique consistent and legend-explicit table along with hard-coded links to the full resolution photos (stored on the GISCO Platform, \url{https://gisco-services.ec.europa.eu/lucas/photos/}).
The LUCAS primary data includes alpha-numerical variables and photographs linked to the geo-referenced points.

% \usepackage{booktabs}
\begin{table}[!!!h]
\footnotesize
\caption{Aggregation of micro data - summarizing the different steps applied to harmonise the data.}
\label{tab:methodMicroData}

\begin{tabular}{@{}m{1.1cm}m{1cm}m{1cm}m{1cm}m{2.4cm}m{3cm}m{0.7cm}@{}}
\hline
\textbf{Source} & \textbf{Survey} & \textbf{Points (n)} & \textbf{Year} & \textbf{Protocol 1} & \textbf{Protocol 2} & \textbf{Data} \\ \hline
\cite{l2006BE,l2006CZ,l2006DE,l2006ES,l2006FR,l2006IT,l2006LU,l2006HU,l2006NL,l2006PL,l2006SK} & Micro data survey 2006 & \num{168401} & 2006 & Data download, Data documentation \cite{cd2006,c12006}, Preparation (year aggregation), Generate mapping files  & Column renaming (Table \ref{tab:changeName}) , Missing column adding (Table \ref{tab:addCols}), New column adding (Table \ref{tab:new_cols}), Character case uniformity, Variable re-coding (Table \ref{tab:recoding}), Column order & Data 2006 \\
\hline
\cite{l2009} & Micro data survey 2009 & \num{234623} & 2009 & Data download, Data documentation \cite{cd2009,c12009}, Preparation, Generate mapping files  & Column renaming (Table \ref{tab:changeName}), Missing column adding (Table \ref{tab:addCols}), New column adding (Table \ref{tab:new_cols}), Character case uniformity, Variable re-coding (Table \ref{tab:recoding}), Column order & Data 2009  \\
\hline
\cite{l2012} & Micro data survey 2012 & \num{270272} & 2012 & Data download, Data documentation \cite{cd2012,c12012}, Preparation, Generate mapping files & Column renaming (Table \ref{tab:changeName}), Missing column adding (Table \ref{tab:addCols}), New column adding (Table \ref{tab:new_cols}), Character case uniformity, Variable re-coding (Table \ref{tab:recoding}), Column order & Data 2012 \\
\hline
\cite{l2015} & Micro data survey 2015 & \num{340143} & 2015 & Data download, Data documentation \cite{rd2015}, Preparation, Generate mapping files  &  Column renaming (Table \ref{tab:changeName}), Missing column adding (Table \ref{tab:addCols}), New column adding (Table \ref{tab:new_cols}), Character case uniformity, Variable re-coding (Table \ref{tab:recoding}), Column order & Data 2015  \\
\hline
\cite{l2018} & Micro data survey 2018 & \num{337854} & 2018 & Data download, Data documentation \cite{{rd2018}}, Preparation, Generate mapping files &  Missing column adding (Table \ref{tab:addCols}), New column adding (Table \ref{tab:new_cols}), Character case uniformity & Data 2018 \\ \hline
\end{tabular}
\end{table}

%Tables should be used to support the data records, and should clearly indicate the samples and subjects (study inputs), their provenance, and the experimental manipulations performed on each (please see Tables and Submitting Experimental Metadata, below). They should also specify the data output resulting from each data-collection or analytical step, should these form part of the archived record.

\subsection{Micro data collection and documentation (Protocol 1)}
\label{sec:protocol1}

The first step is to collect the data from the source for each survey year (see Table \ref{tab:methodMicroData} Source).
The raw micro data for the harmonised database presented here are the five LUCAS campaigns, which can be downloaded from the official Eurostat website \cite{estatLucasSite}. The LUCAS micro data for 2006 is divided into a separate file for each of the 11 surveyed countries (Belgium \cite{l2006BE}, Czechia \cite{l2006CZ}, Germany \cite{l2006DE}, Spain \cite{l2006ES}, France \cite{l2006FR}, Italy \cite{l2006IT}, Luxembourg \cite{l2006LU}, Hungary \cite{l2006HU}, Netherlands \cite{l2006NL}, Poland \cite{l2006PL}, and Slovakia \cite{l2006SK}). The LUCAS micro data from the other survey years is served in a yearly aggregated  form, whereby the data from all participating countries can be downloaded separately (2009 \cite{l2009}, 2012 \cite{l2012}, 2015 \cite{l2015}) in CSV format and 2018 \cite{l2018} in 7z zipped format. %Along with the survey data, the providershttps://www.overleaf.com/1576786112dmcwdyjcyssb also include the pictures from each of the surveyed points, available for bluk free-of-charge download. 

The second step is to collect the documentation that facilitates translating the alpha-numerical class-description in the original datasets into explicit information. 
For 2006, 2009 and 2012, the survey data comes with a content descriptor (2006 \cite{cd2006}, 2009 \cite{cd2009}, 2012 \cite{cd2012}), though the content descriptor doesn't necessarily have the same number of variables as the data; and the variables themselves sometimes have a slightly different name. These inconsistencies were resolved with assistance from the technical documents (LC1 (Instructions, 2006 \cite{c12006}, 2009 \cite{c12009}, 2012 \cite{c12012}) and LC3 (Classification, 2006 \cite{c32006}, 2009 \cite{c32009}, 2012 \cite{c32012}).
From 2015 and 2018, the data is served with a record descriptor (2015 \cite{rd2015}, 2018 \cite{rd2018}), which provides information on variable name, data type and description, in a more consolidated fashion, making it easier to find information about the relevant variable.\\
The final step in Protocol 1 is the generation of the mapping files, i.e. the schema mapping used in database design and architecture. Due to the nature of the task, instead of performing schema-level matchers, the study fit more within the semantic scope of instance-level matchers, whereby more importance is placed on the contents and meaning of the schema elements. The workflow thus mapped the ascertained relationship between variables that have changed in name or alpha-coding between the surveys. These mappings serve as a blueprint for the transformation and data integration described in Protocol 2.

\subsection{Yearly micro data harmonisation  (Protocol 2)}
\label{sec:protocol2}

\begin{figure}[H]
    \centering
    %\hspace*{-3.7cm}   
    \includegraphics[width=1\textwidth]{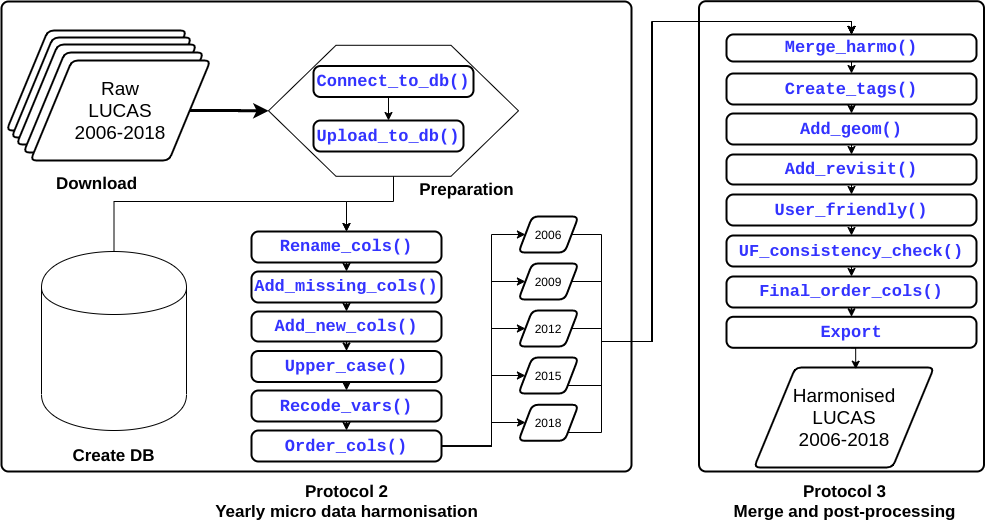}
    \caption{Processing workflow to harmonise the survey data.}
    \label{fig:workflow}
\end{figure}

The harmonisation workflow, alongside the performed database consistency checks, are schematically shown in Figure \ref{fig:workflow} and the code is described in code section (Section \ref{sec:code}). The general principle of the harmonisation workflow was to convert all the field legends to fit with the latest i.e. the 2018 database layout (the next LUCAS is currently planned for 2022). 

Some notable changes in the source tables had to be made in order to make the harmonisation and subsequent merger into one complete table possible. This was accomplished with the above-mentioned instance-mapping files (Section (\ref{sec:protocol1})). All manipulations executed over the separate tables prior to the merger are listed in Table \ref{tab:methodMicroData} under heading 'Protocol 2':
\begin{itemize}
\item Column renaming - iteratively renaming columns to align them with the last (in this case 2018) survey. Performed on all tables but 2018 by using the \textcolor{blue}{\texttt{Rename\_cols()}} function from the package.
\item Missing column adding - iteratively adding all columns that are present in one table and not present in the others. Performed on all tables by using the \textcolor{blue}{\texttt{Add\_missing\_cols()}} function.
\item New column adding - iteratively adding all newly created columns. These include the variables 'letter group', 'year', and 'file\_path\_gisco\_n/s/e/w/p' (for more information check Table \ref{tab:descriptor}). Performed on all tables using the \textcolor{blue}{\texttt{Add\_new\_cols()}} function.
\item Character case uniformity - iteratively converting all characters of selected fields to upper case. Performed on all tables using the \textcolor{blue}{\texttt{Upper\_case()}} function.
\item Variable re-coding - iteratively re-coding selected variables according to created mapping CSV files, designed by consultation with reference documents. Performed on all tables but 2018 by using the \textcolor{blue}{\texttt{Recode\_vars()}} function.
\item Column order - iteratively ordering all columns according to the template from the 2018 survey. Performed on all tables but 2018 by using the \textcolor{blue}{\texttt{Column\_order()}} function.

\end{itemize}

%These had primarily do to with changing the coding of the variables, introducing new codings in order not to lose information for the sake of harmonisation (as is the case with the variable obs\_type where an extra value, corresponding to Marine sea was added), introducing new variables altogether (Table \ref{tab:new_cols}), and renaming columns to align them with the 2018 survey (Table \ref{tab:changeName}). Another essential part of the harmonisation was adding all relevant columns to the separate primary data tables, populating with null values where necessary, and arranging them in the same order before joining them. 

The full workflow is  dependent on two software prerequisites. Firstly, one must have a running PostgreSQL server, and secondly, an installation of R (more about the versions used in section \ref{sec:code}). The pipeline is provided as a R package for ease of reproducability and transparency (section \ref{sec:code}).

\subsection{Merge and post-processing (Protocol 3)}

The third part of the harmonisation process includes the merging of the yearly harmonised tables plus additional steps listed below before exporting the final data outputs.

\begin{itemize}

\item Merge yearly harmonised table - merge the 5 harmonised tables to one unique table via \textcolor{blue}{\texttt{Merge\_harmo()}} function.

\item Create database primary key, index, and spatial index via the  \textcolor{blue}{\texttt{Create\_tags()}} function.

\item Add geometries and calculated distances - location of theoretical point(\textit{th\_geom}), location of lucas survey (\textit{gps\_geom}), lucas transect geometries (\textit{trans\_geom}) and distance between theoretical and survey point (\textit{th\_gps\_dist}). Done by the  \textcolor{blue}{\texttt{Add\_geom()}} function.

\item Add revisit column to lucas harmonized table to show the number of times between the years when the point was revisited thanks to the \textcolor{blue}{\texttt{Add\_revisit()}} function.

\item Create columns with labels for coded variables and decodes all variables where possible to explicit labels. Performed with \textcolor{blue}{\texttt{User\_friendly()}} function.

\item Perform consistency checks on newly created fields to ensure conformity in terms of column order and data types via the \textcolor{blue}{\texttt{UF\_Consistency\_check()}} function.

\item Re-order columns of final tables with the \textcolor{blue}{\texttt{Final\_order\_cols()}} function.

\item Export the outputs. The table is exported as CSV and the geometries as shapefiles.

\end{itemize}

\subsection{Full resolution LUCAS photos}
In addition to the alphanumerical and geometry information of the survey, a complete database with full-resolution point and landscape photos was designed with photos retrieved from Eurostat. This archive was organized as a table with all the exchangeable image file (EXIF) variables for each of the images, among which a unique file path, as stored on the Eurostat GISCO cloud service for easy retrieval by other researchers. Besides the EXIF attributes, each photo is also hard-coded with the respective point ID of the LUCAS point and the year of survey. The photos' metadata were extracted with ExifTool (v 10.8) \cite{harvey2013exiftool} resulting in a database of photos that was compared for completeness with the survey data records. The hard-coded HTTPS links to each photo in the consolidated database allow for massive query and selection tasks. 

\section{Data Records}
\label{sec:dataRecords}

%The Data Records section should be used to explain each data record associated with this work, including the repository where this information is stored, and to provide an overview of the data files and their formats. Each external data record should be cited numerically in the text of this section, for example \cite{cite3, cite4, cite5, cite6}, and included in the main reference list as described below.. A data citation should also be placed in the subsection of the Methods containing the data-collection or analytical procedure(s) used to derive the corresponding record.

\subsection{Storage}

\begin{itemize}
  \item \textbf{Multi-year harmonised LUCAS survey data.} The harmonised database (available here \url{https://figshare.com/s/4a2e5d119ee0a98bec6e} \cite{lucasHarmoFigshare}) contains 117 variables and 1,351,293 records corresponding to a unique combination of year and location. The same dataset is also available for each year with a different file for users interested only in one specific survey. The database is served with a \textbf{Record descriptor} (Table \ref{tab:descriptor} presents a summary, the complete table is available here in CSV format \cite{lucasHarmoFigshare}). This record descriptor specifies variable name, data type, range of possible values and meanings. In the documentation one can find more information about the variable and a short description, along with comments concerning the variable that the authors have deemed important. Additionally, the tables in  \textbf{LUCAS-Variable and Classification Changes} provide documentation for users to quickly identify the differences between yearly LUCAS campaigns and the harmonised database. The file contains four tables:
  \subitem
	1) “References” : Description and a legend of the used colors of the different tables;
	\subitem
	2) “Harmonised DB”: a comparison of all the collected variables of the 2018 survey with the variables of the harmonised database and an overview of the actions to harmonize the data;
	\subitem
 	3) “Variable changes”: an overview/ comparison of all collected variables between all campaigns from 2009 to 2018 highlighting the changes;
 	\subitem
	4) “LC (LU) changes” an overview of the possible LC and LU codes of each campaign highlighting the changes.
  \item \textbf{High resolution LUCAS photo archive.} The 5.4 millions of photos collected during the five surveys are available here: \url{https://gisco-services.ec.europa.eu/lucas/photos/}. For each in-situ point, landscape (N, E, S, W), and point (P) photos are available. 
  \item \textbf{LUCAS survey geometries/point locations.} To facilitate spatial analysis and usability, three types of geometries are provided as distinct shapefiles (\url{https://figshare.com/s/4a2e5d119ee0a98bec6e}) : 
  \subitem 
  (i) LUCAS theoretical points (\textit{th\_long}, \textit{th\_lat}),
  \subitem
  (ii) LUCAS observed points (\textit{gps\_lon}, \textit{gps\_lat}) and
  \subitem
  (iii) LUCAS transect lines (250-m east looking lines).
  \item \textbf{A R package.} The scripts to harmonise the LUCAS data is provided as an open source R package (downloadable here for review \url{https://www.dropbox.com/sh/dwhoj1p4izds9bh/AABckSM_zNLZxzy613Yc5lIFa?dl=0})% and will be later pushed to the CRAN)). %(\url{https://cran.r-project.org/web/packages/lucas/}) along with the documentation (\url{https://cran.r-project.org/web/packages/lucas/lucas.pdf}).
\end{itemize}

\subsection{Overview of multi-year harmonised LUCAS survey database}
\label{sec:o_survey}
During five LUCAS surveys, a total of \num{1351293} observations have been made at \num{651676} unique locations (Table \ref{tab:pointsPerCountry}). The total number of surveyed points has increased significantly from the 2006 pilot study (\num{168401}) to 2015 (\num{340143}) (Table \ref{tab:pointsPerCountry}). This rise is mainly due to the increase in terms of thematic richness, scope, and scale of the study from what was primarily an evaluation of agricultural areas (2006) to a more holistic and exhaustive inspection of the EU territory. Additional, the total number of countries surveyed increased from 11 in 2006 to 28 in 2018 (Table \ref{tab:pointsPerCountry}). Over the five surveys, \num{980688} observations (73\%) were done in-situ, i.e. either 'In situ < 100 m', 'In situ > 100 m').
The proportion of points where actual in-situ data was collected has decreased from 92\% in 2006 to 60\% in 2018. Furthermore 11\% of the points (i.e. \num{147574}) that were visited in-situ turned out not be accessible in practice and are photo-interpreted in the field. The number of points surveyed per country and per year ranged between 79 (Malta) to \num{48215} (France). Finally, over the five surveys, 1677 points were out of national territory, i.e. "NOT EU" corresponding to water outside national border or countries including Russia, Turkey, Albania and Switzerland).

\setlength{\tabcolsep}{3pt}

\begin{table}[!h]
\caption{LUCAS points per country and per year. The total number of records is provided by year and also split according to the type of observation: In-situ (direct observation), In-situ PI (In-situ Photo-Interpreted if point is not accessible) or Office PI (Photo-interpreted in the office and thus not in-situ)} 
\label{tab:pointsPerCountry}
\footnotesize
\centering
\sisetup{tight-spacing=true,table-number-alignment=center}

%\begin{tabular}{>{\bfseries} c | *{5}{S} | S }
\begin{tabular}{>{\bfseries} c | 
S[table-column-width=0.6cm]
S[table-column-width=0.6cm]
S[table-column-width=0.6cm]
S[table-column-width=0.6cm]
S[table-column-width=0.6cm]|
S[table-column-width=0.6cm] }
  \hline
 \textbf{ } & \textbf{2006} & \textbf{2009} & \textbf{2012} & \textbf{2015} & \textbf{2018} & \textbf{Total \# records} \\ 
  \hline
 AT & 0 & 4961 & 6469 & 8839 & 8840 & 29109 \\
  BE & 2370 & 1804 & 2446 & 2899 & 3659 & 13178 \\
  BG & 0 & 0 & 6641 & 7677 & 7678 & 21996 \\
  CY & 0 & 0 & 1442 & 1726 & 2313 & 5481 \\
  CZ & 5626 & 4662 & 5514 & 5712 & 5713 & 27227 \\
  DE & 27507 & 21113 & 24939 & 26598 & 26777 & 126934 \\
  DK & 0 & 2540 & 3442 & 3665 & 3703 & 13350 \\
  EE & 0 & 2663 & 2200 & 2637 & 2665 & 10165 \\
  EL & 0 & 7758 & 7821 & 12521 & 12622 & 40722 \\
  ES & 34489 & 29912 & 35377 & 50281 & 45314 & 195373 \\
  FI & 0 & 19895 & 13476 & 16116 & 16182 & 65669 \\
  FR & 39070 & 32318 & 38324 & 48188 & 48215 & 206115 \\
  HR & 0 & 0 & 0 & 3532 & 4239 & 7771 \\
  HU & 8422 & 5513 & 4637 & 5169 & 5514 & 29255 \\
  IE & 0 & 4164 & 3484 & 4907 & 4975 & 17530 \\
  IT & 20291 & 17790 & 20985 & 28693 & 28294 & 116053 \\
  LT & 0 & 3860 & 3889 & 4505 & 4584 & 16838 \\
  LU & 197 & 152 & 213 & 251 & 340 & 1153 \\
  LV & 0 & 3825 & 4420 & 5374 & 5376 & 18995 \\
  MT & 0 & 0 & 79 & 79 & 79 & 237 \\
  NL & 2916 & 2449 & 2237 & 2521 & 5011 & 15134 \\
  PL & 24128 & 18487 & 21797 & 22980 & 23086 & 110478 \\
  PT & 0 & 5423 & 7332 & 9006 & 7168 & 28929 \\
  RO & 0 & 0 & 14278 & 16720 & 16725 & 47723 \\
  SE & 0 & 26656 & 22420 & 26648 & 26709 & 102433 \\
  SI & 0 & 1203 & 1621 & 1923 & 1922 & 6669 \\
  SK & 3385 & 2898 & 2455 & 2755 & 2898 & 14391 \\
  UK & 0 & 14438 & 12214 & 16803 & 17253 & 60708 \\
 NOT EU & 0 & 139 & 120 & 1418 & 0 & 1677 \\
 \hline
  Total \# records & 168401 & 234623 & 270272 & 340143 & 337854 & 1351293 \\
  Total \# countries & 11 & 23 & 27 & 28 & 28 &  \\
 \hline
  In-situ  \# & 155238 & 161674 & 229004 & 230811 & 203961 & 980688 \\
  In-situ  [\%] & 92 & 69 & 85 & 68 & 60 & 73 \\
  In-situ PI \# & 13163 & 59594 & 26669 & 25254 & 22894 & 147574 \\
  In-situ PI [\%] & 8 & 25 & 10 & 7 & 7 & 11 \\
  Office PI \# & 0 & 13355 & 14599 & 83982 & 110962 & 222898 \\
  Office PI [\%] & 0 & 6 & 5 & 25 & 33 & 16 \\

  \hline
%\end{tabulary}
\end{tabular}

\end{table}

\begin{figure}[!h]
    \centering
    \includegraphics[width=\textwidth]{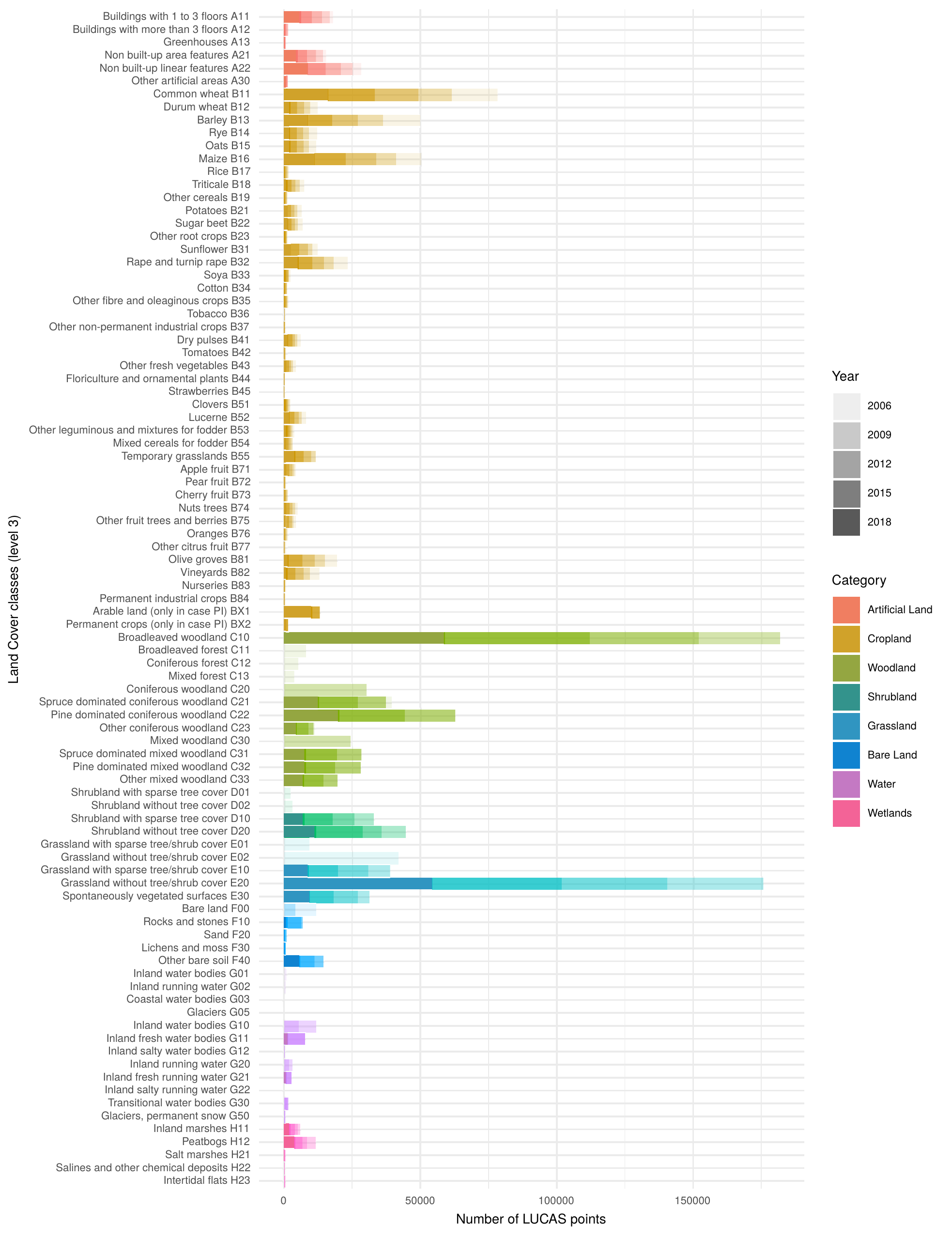}
    \caption{Distribution of land cover classes in the multi-year harmonised LUCAS survey.}
    \label{fig:LUCAS_harmo_LC1_level3_Distribution}
\end{figure}

Figure \ref{fig:LUCAS_harmo_LC1_level3_Distribution} shows the distribution of the surveyed points in 77 detailed legend level-3 classes by year. Land Cover/Land Use (LC/LU) classification specifications can be found in reference document C3 (\cite{c32018}). The classification system follows rules on spatial and temporal consistency - it can be applied and compared both between locations in the EU, and years. Additionally, it is compatible with other existing LC/LU systems (e.g. Food and Agriculture Organization (FAO), statistical classification of economic activities in the European Community (NACE)) and fulfills the specifications of the European Infrastructure for Spatial Information in Europe (INSPIRE) standardization initiative on LC/LU. To inform about changes in two consecutive surveys, the data providers describe the adjustments to the terminology in the documentation.   
The 3-level legend system is stored hierarchically, whereby the first level (letter group) corresponds to the eight main classes obtained by ortho-photo-interpretation during the second level stratification phase (figure \ref{fig:harmo}); the second and third level, representing subcategories of these main classes are indicated by a combination of the letter group and further digits. 

\begin{figure}[!!!h]
    \centering
    \includegraphics[width=0.95\textwidth]{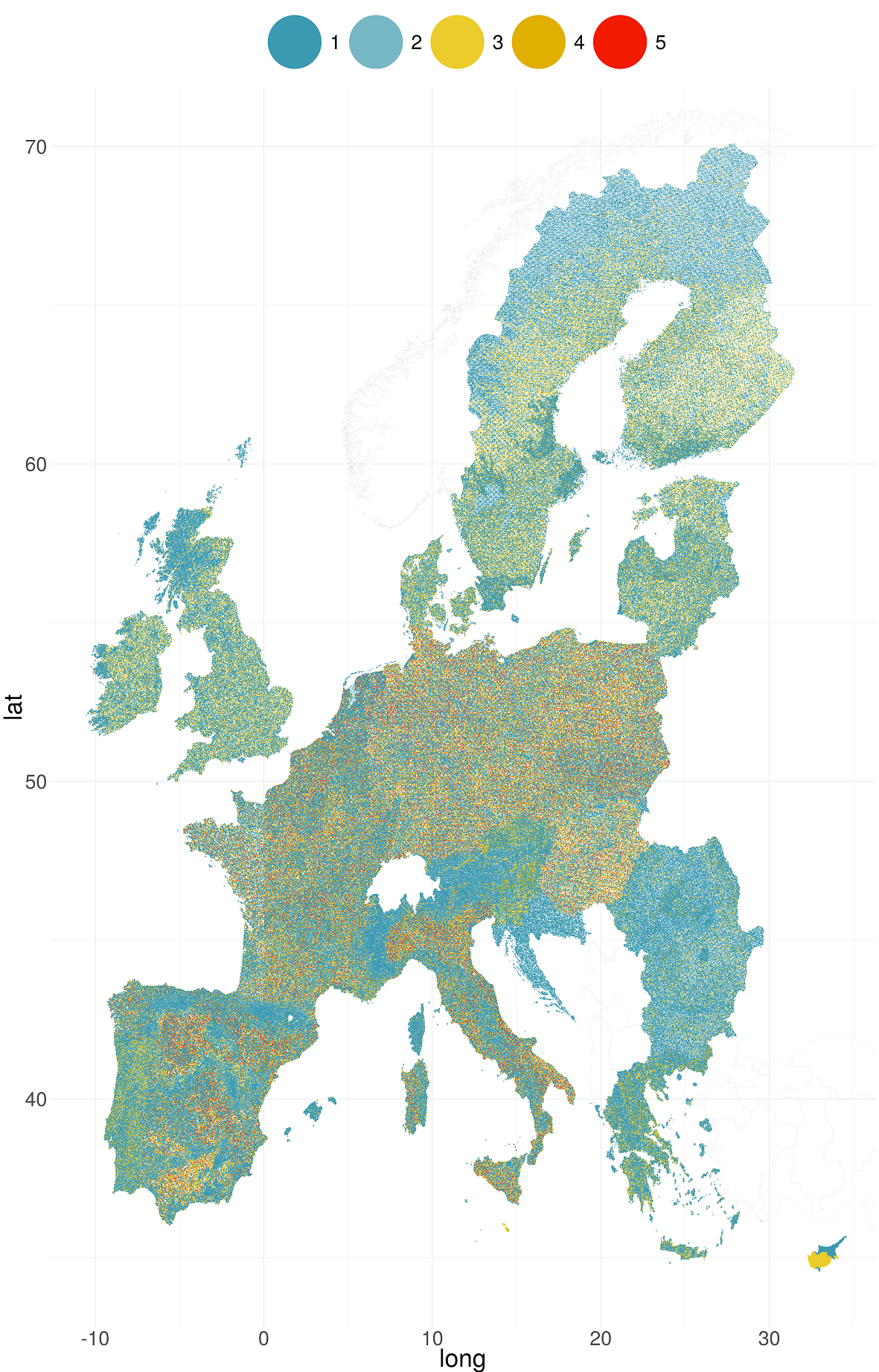}
    \caption{Over the five surveys between 2006 and 2018, \num{651780} points were at least surveyed once. Revisit ranges from one to five. }
    \label{fig:lucas_harmo_revisit}
\end{figure}

The number of point revisits is shown in Table \ref{tab:lucas_harmo_revisit_number}. Some LUCAS points were visited once in 15 years (n=\num{332605}) while others were visited each time, thus totaling five revisits (n=\num{35204}). This means that \num{651780} locations were at least visited once. Figure \ref{fig:lucas_harmo_revisit} shows a map with the revisit frequency for each point over Europe.

\begin{table}[!h]
\centering
\footnotesize
\sisetup{tight-spacing=true,table-number-alignment=center}
\caption{Number of LUCAS points and revisits}
\begin{tabular}{C{4cm} 
S[table-column-width=1.2cm]
S[table-column-width=1.2cm]
S[table-column-width=1.2cm]
S[table-column-width=1.2cm]
S[table-column-width=1.2cm]
}
  \hline
\textbf{Revisit} & 1 & 2 & 3 & 4 & 5 \\ 
   \hline
\textbf{Number of points} & 332605 & 101052 &  91112 &  91807 &  35204 \\
   \hline
\end{tabular}
\label{tab:lucas_harmo_revisit_number}
\end{table}

\subsection{Overview of EXIF photos database}
\label{sec:o_pictures}

The available photos (N, E, S, P) were cataloged totaling \num{5440459} photos for the 5 surveys (see Table \ref{tab:exif} for detailed distribution).
While the observation location is recorded by the surveyor in the LUCAS survey (\textit{gps\_lon}, \textit{gps\_lat}), the digital cameras with GPS could also capture the location where the photos were taken as well as the direction, \textit{i.e.} the azimuth angle. In the first surveys, the digital camera and the GPS were separate devices. The direction was determined with a traditional compass. The data were used to cross-validate the geo-location reported during the survey. 
To assess the availability of this information, the EXIF information of the \num{5440459} photos was retrieved. As summarised in the two last columns of Table \ref{tab:exif}, the photos with geo-location information have increased considerably through time, i.e. 0\% in 2006, 5.4 \% in 2009, 34.2.8\% in 2012, 68.5\% in 2015 and finally 72.9\% in 2018. For azimuth angle, there is no information on orientation for the photos taken in 2006 and 2009. However, respectively 15.3\%, 22\%,and 6.7\% of the photos have EXIF direction information for 2012, 2015, and 2018.

\begin{table}[!h]
\centering
\caption{Number of LUCAS photos per year, per type (N, E, S, W, P) proportions that have EXIF geo-location (Location [\%])  and orientation information (Orientation [\%])}
\label{tab:exif}
\sisetup{tight-spacing=true,table-number-alignment=center,table-format=7.0}
\footnotesize

\begin{tabular}{>{\bfseries}C{1cm} |
S[table-column-width=0.8cm]
S[table-column-width=0.8cm]
S[table-column-width=0.8cm]
S[table-column-width=0.8cm]
S[table-column-width=0.9cm] |
S[table-column-width=1cm] |
C{1cm}|C{1cm}}

% \begin{tabular}{@{}m{0.7cm}|m{0.9cm}m{0.9cm}m{0.9cm}m{0.9cm}m{0.9cm}|m{1,3cm}|m{1,3cm}|m{1,3cm}@{}}
  \hline
 \textbf{Year} & \textbf{East} & \textbf{North} & \textbf{Point} & \textbf{South} & \textbf{West} & \textbf{TOTAL} & \textbf{Location [\%]} & \textbf{Orientation [\%]} \\ 
  \hline
  2006 & 137461 & 137426 & 134538 & 137368 & 137179 & 683972 & 0 & 0 \\ 
  2009 & 199208 & 199264 & 171165 & 199129 & 199117 & 967883 & 5.4 & 0 \\ 
  2012 & 269329 & 269286 & 243074 & 269277 & 269205 & 1320171 & 34.2 & 15.3 \\ 
  2015 & 265421 & 265392 & 242772 & 265368 & 265285 & 1304238 & 68.5 & 22 \\ 
  2018 & 237259 & 237529 & 215190 & 237262 & 236955 & 1164195 & 72.9 & 6.7 \\
  \hline
  Total & 1108678 & 1108897 & 1006739 & 1108404 & 1107741 & 5440459 &  &  \\ 
   \hline
\end{tabular}
\end{table}

\begin{figure}[!!h]
    \centering
    \includegraphics[width=\textwidth]{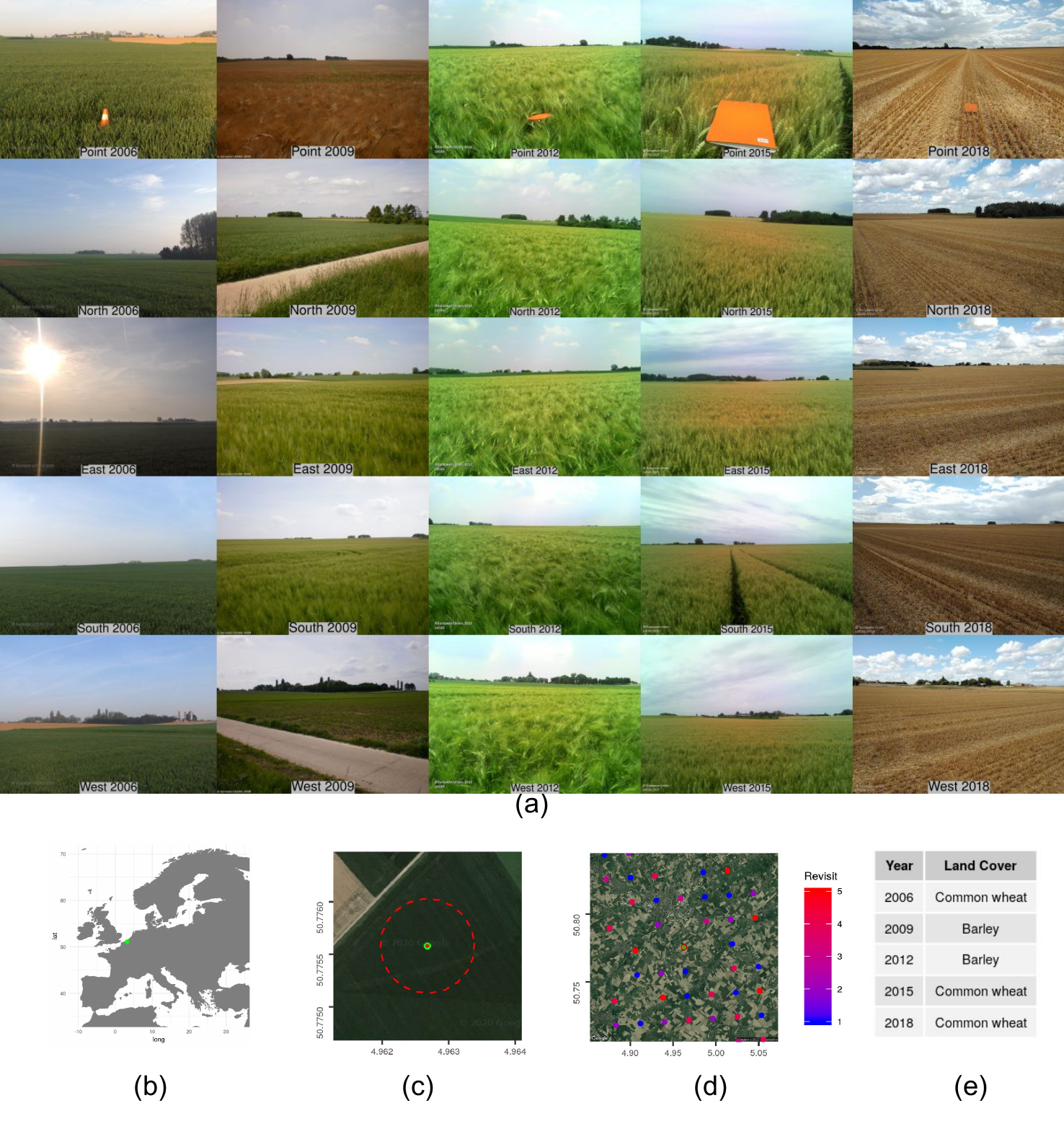}
    \caption{Overview of the data available for a LUCAS point that was revisited five times: (a) Point, North, East, South and West photos for 2006, 2009, 2012, 2015 and 2018, (b) Location of the point in the EU, (c) Zoom showing the point (3-m diameter in green, 50-m diameter in dashed red, (d)  Revisit frequency on a 20 by 20 km square centered on the point, and (e) In-situ land cover observation of the point for the different years.}
    \label{fig:dashboard}
\end{figure}

Each point surveyed has potentially five photos (N, E, S, W, P) per surveyed year (Figure \ref{fig:dashboard} (a)). This part of the database is represented by a table of records, corresponding to the photos taken in the cardinal directions plus the point for each one of the points for the five surveys. The table holds information on the point ID, year of survey, path to the full resolution image and an wide variety of EXIF extracted attributes, among which are coordinates, orientation, camera model, date, Eurostat metadata, etc.

% FigDashBoard.pdf 

It was decided that having this information in a separate table is more sensible in terms of storage and accessibility, whereby cross-table checks can easily be performed by executing joins between the tables based on point ID and year of survey. As such, by combining this information from the two tables together one arrives at a significantly large set of labeled examples, corresponding to images of the 77 different types of recorded land cover - an until-now-untapped resource for e.g. deep learning aided computer vision exercises.

\section{Technical Validation}

%This section presents any experiments or analyses that are needed to support the technical quality of the dataset. This section may be supported by up figures and tables, as needed. This is a required section; authors must present information justifying the reliability of their data.

Observations are subject to detailed quality checks (see LUCAS metadata \cite{LUCASmetadata} and the data quality control documents available for 2009 \cite{c42009}, 2012 \cite{c42012}, 2015 \cite{c42015}). First, an automated quality check verifies the completeness and consistency after field collection. Second, all surveyed points are checked visually at the offices responsible for collection. Third, an independent quality controller  interactively checks 33\% of the points for accuracy and compliance against pre-defined quality requirements, including the first 20\% observations for each surveyor, to prevent systematic errors during the early collection phase.  

The present data consolidation effort seeks to enhance the quality of an already existing product. Ensuring data quality while harmonising through years is thus essential. %as not doing so would have detrimental effects on decision-making, funds allocation, and scientific discovery. 
%Moreover,  %As such the importance of upholding the established data quality standards cannot be stressed enough. 
Data quality was ensured by taking account of validity, accuracy, completeness, consistency, and uniformity throughout  data processing (Figure \ref{fig:workflow}). In the first step, the surveys were harmonised, in a second step the resulting database was made user-friendly by incorporating legend-explicit values. 

Validity was ensured via data type (for which information can be found in the record descriptor) and a unique constraint of a composite key (consisting of the point ID and year of survey). The accuracy of the data relies on the source data. Completeness checking reveals that since several variables have been added over the years, many missing values exist. In such cases, fields were populated with null values. Consistency across surveys has been greatly enhanced. All surveys were harmonised towards the 2018 survey. Internal data consistency of the presented data set was ensured through running checks at various stages of processing. Uniformity checks revealed that the  geographical coordinates in columns \textit{th\_long} and \textit{th\_lat} show different locations between some years. In the interest of complete uniformity, it was decided to have the values of these variables hard coded from the LUCAS grid. Because the LUCAS grid is a non-changing and set in stone feature of the LUCAS survey, the locations of each point remains the same throughout the years. Thus any discrepancy between the recorded theoretical location of a LUCAS point in the micro data and the grid must be corrected. This was done for all but 64 points from 2006 which are not in the grid.

%for the 2006 survey, were given in Lambert Azimuthal Equal Area (LAEA, EPSG:3035) projection. These were reprojected to geographical long, lat coordinates (EPSG:4326) to match with the other survey data. 

To further asses spatial accuracy of the data, the distance between the theoretical point from the LUCAS grid (\textit{th\_long}, \textit{th\_lat}), and the actual GPS measurement of the survey observation point (\textit{gps\_lon},\textit{gps\_lat}) were compared. 
This is important for several reasons - firstly, it allows to ascertain the real distance between the point actually surveyed and the point supposed to be surveyed, which is, in a sense, a proxy for the quality of the surveyed observation itself; secondly, it is an accuracy check of the surveyed distance between the theoretical point and the survey observation point, as collected by the surveyor, "as provided by the GPS (in m)" (column \textit{obs\_dist}), and the distance between the same points as calculated from the data (column \textit{th\_gps\_dist}). It must be noted that for the 2006 survey the variable \textit{obs\_dist} was recorded as a range, whereas for the other years it represents the actual value of the distance. For this reason it was decided to hard code the values for 2006 to match exactly with the calculated distance. In this way we ensure consistency in the data type of the column, yet sacrifice the nuances from changing the original data. Table \ref{tab:th_dist_TF} gives a breakdown of this, whereby in 2006 we see the 100\% match between recorded and calculated distance because of the reason just explained, for 2009 a match of 96.3\%, meaning that for only 3.7\% of the cases did the value not match.

%86.59\% of the points have a recorded value of distance between points (column \textit{obs\_dist}), which matches the value calculated, and consequently 13.41\% of points for which these distances do not match. 

In carrying out this comparison it became apparent that the percent of matching distances has increased through years probably due to better precision of positioning sensors; the total amount of error in 2018 is reduced to a negligible 0.31\%; and lastly, that there were flagged and removed a number of records that have inaccurate GPS coordinates most probably due to sensor malfunction. Cross-checking with the source data, we found that the error is indeed present in the source data, rather than introduced during processing - something which would have been hard to spot otherwise.

% latex table generated in R 3.4.3 by xtable 1.8-2 package
% Thu Oct 24 12:31:53 2019
% \begin{table}[ht]
% \centering
%   \caption{Percentage (\%) of points for which the distances between the theoretical point from LUCAS grid (\textit{th\_long}, \textit{th\_lat}) and the actual GPS  observation measurement ({gps\_lon},\textit{gps\_lat}) taken during surveying and calculated post factum match or not.}
%   \label{tab:th_dist_TF}
% \begin{tabular}{rrrrrr}
%   \hline
%  & 2006 & 2009 & 2012 & 2015 & 2018 \\ 
%   \hline
% \textbf{Match} & 86.59 & 96.33 & 97.74 & 99.08 & 99.69 \\ 
%   \textbf{No match} & 13.41 & 3.67 & 2.26 & 0.92 & 0.31 \\ 
%   \hline
% \end{tabular}
% \end{table}

\begin{table}[ht]
\centering
  \caption{Percentage (\%) of points for which the distances between the theoretical point from LUCAS grid (\textit{th\_long}, \textit{th\_lat}) and the actual GPS  observation measurement ({gps\_lon},\textit{gps\_lat}) taken during surveying and calculated post factum match or not.}
  \label{tab:th_dist_TF}
\begin{tabular}{rrrrrr}
  \hline
 & 2006 & 2009 & 2012 & 2015 & 2018 \\ 
  \hline
Match & 100.00 & 96.32 & 97.92 & 99.08 & 99.77 \\ 
  No match & 0.00 & 3.68 & 2.08 & 0.92 & 0.23 \\ 
   \hline
\end{tabular}
\end{table}

%   Min.  1st Qu.   Median     Mean  3rd Qu.     Max. 
%   0.00     1.13     2.78   124.14    21.01 99844.19 

\begin{figure}[!!h]
  \centering
  \begin{subfigure}{.47\linewidth}
    \centering
    \includegraphics[width = \linewidth]{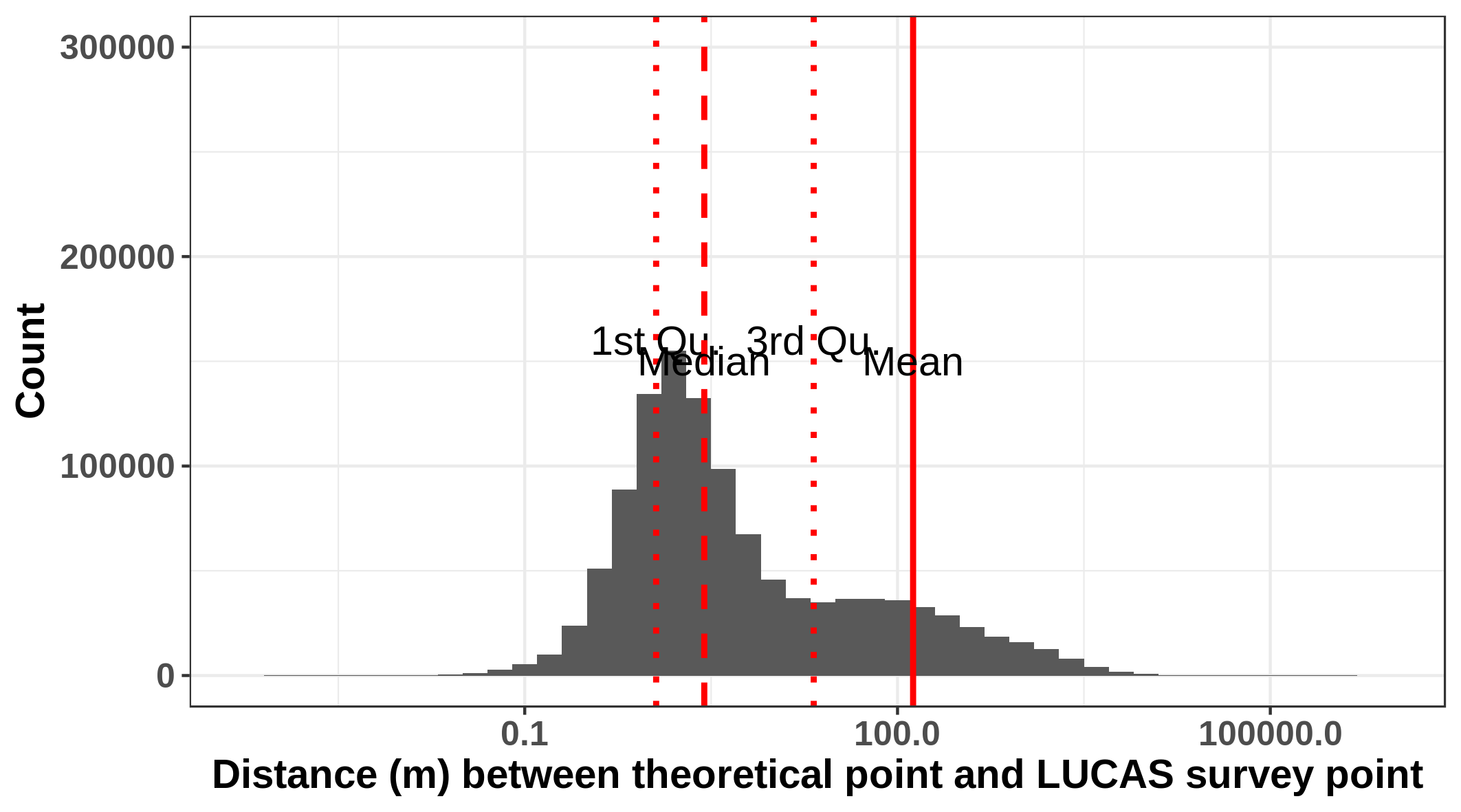}
    \caption{Calculated distances}
  \end{subfigure}
  \hspace{1em}
  \begin{subfigure}{.47\linewidth}
    \centering
    \includegraphics[width = \linewidth]{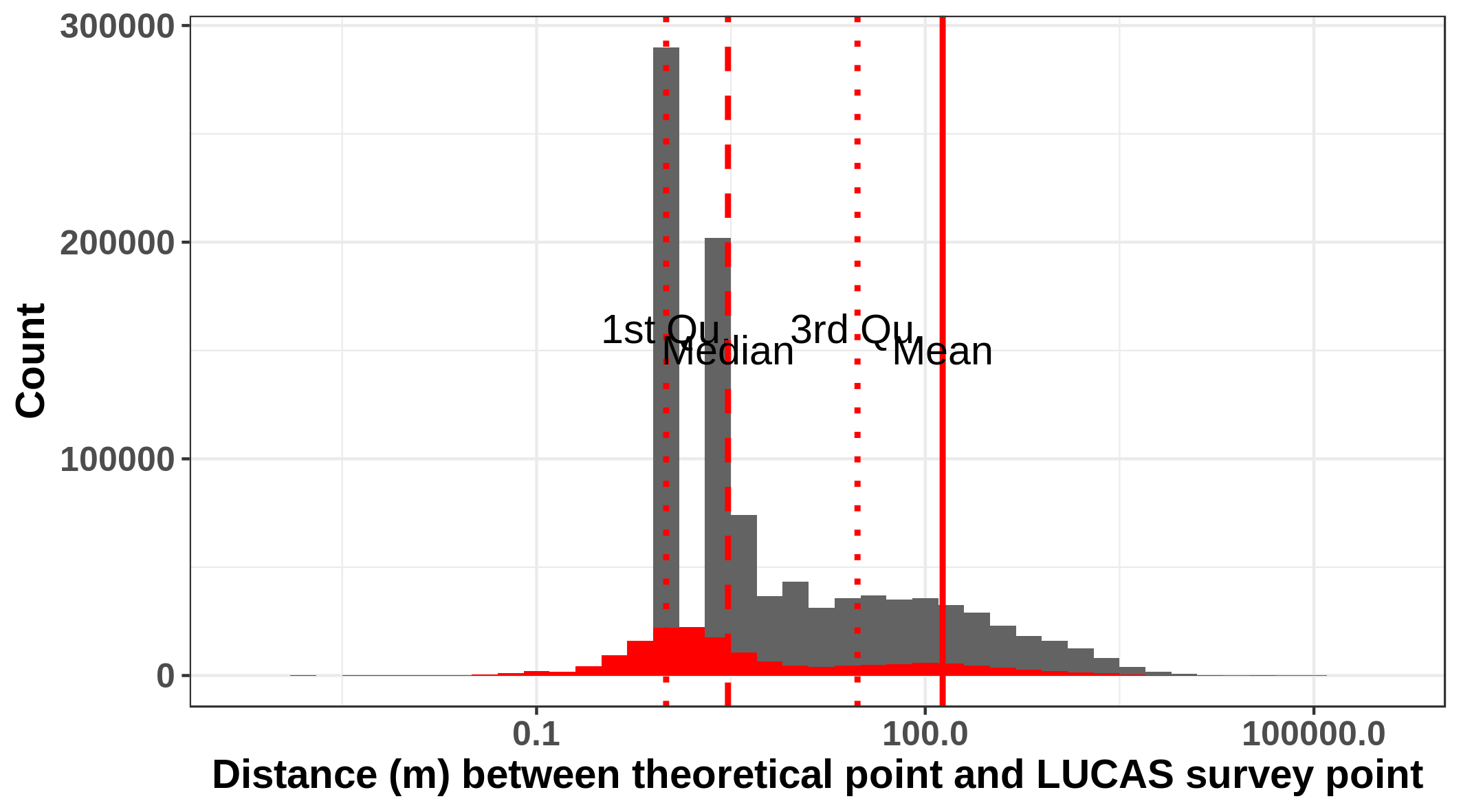}
    \caption{Surveyed distances}
  \end{subfigure}
\caption{Comparison of distributions between calculated and surveyed distances between LUCAS Theoretical Points and Observation Points. The red-colored part of the distribution in subfigure b) represents the data from 2006, which is copied from the calculated distances (\textit{th\_gps\_dist}).}
\label{fig:LUCAS_histo_th_gps_dist}  
\end{figure}

The distribution of these calculated distances, alongside an equivalent distribution of the surveyed distances, can be found in Figure \ref{fig:LUCAS_histo_th_gps_dist}. The distance between 75\% of the points (1-3 quantile) is between 1.1 and 21.2 meters, meaning that  only a fourth of the points have a distance greater than this. For the surveyed distances the ranges are similar - 75\% of the values fall between 1.0 and 30.0 meters. From the distributions we see that there is a lot more nuance in the values of the calculated distances, which makes sense as they are represented by numbers with decimals, which have a lower frequency than the integers, representing the surveyed distances. The values shown in the red part of the histogram of surveyed distances represents the values from 2006, which are copied from the calculated distance in order to hard code a numerical in the place of the otherwise categorical value of the variable in the source data.

\begin{figure}[!!h]
  \centering
  \begin{subfigure}{.47\linewidth}
    \vspace{0.1cm}
    \centering
    \includegraphics[width = \linewidth]{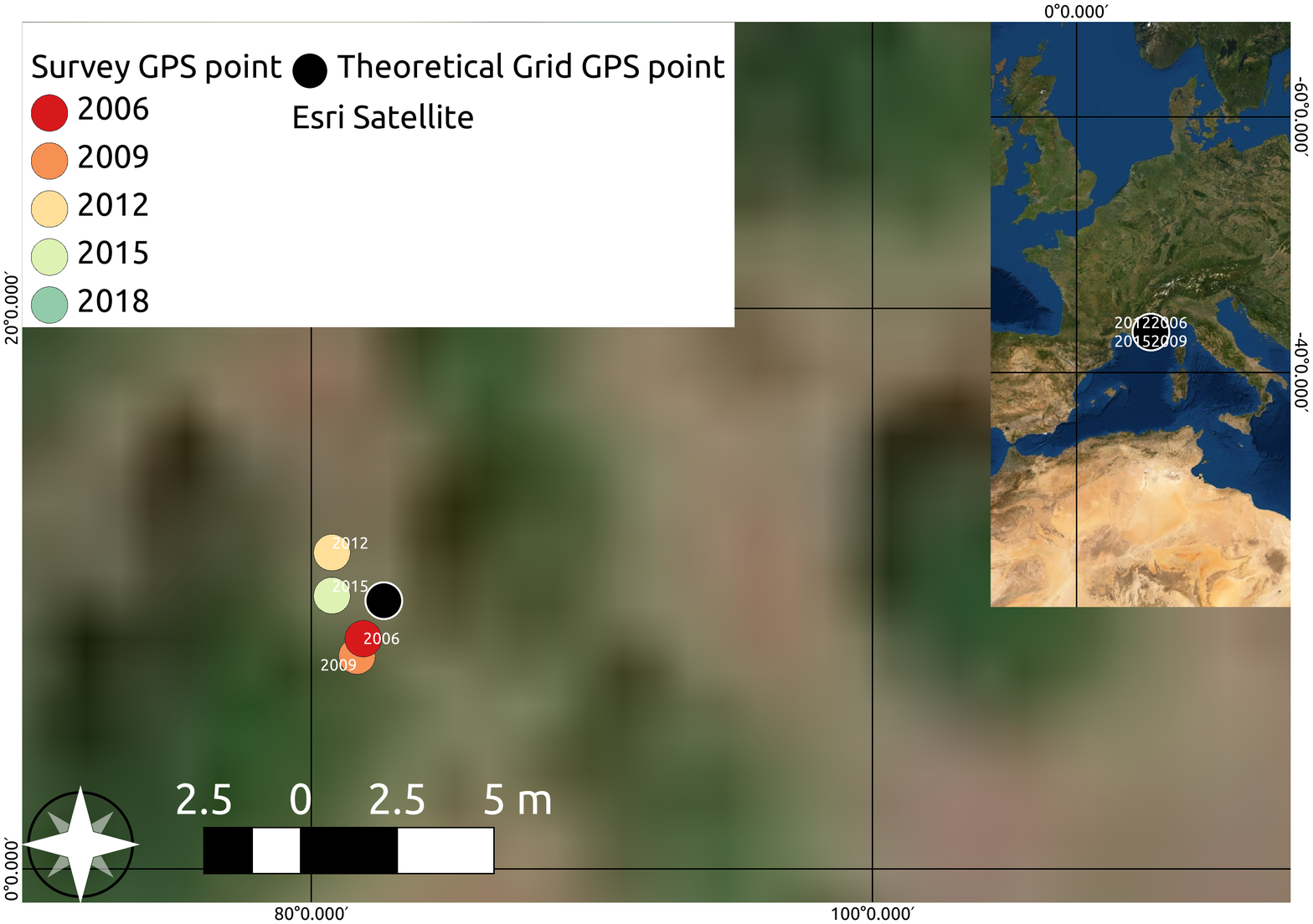}
    \caption{Example of a surveyed point (id 40402278) at close distance (< 2m)}
  \end{subfigure}
  \hspace{1em}
  \begin{subfigure}{.47\linewidth}
    \centering
    \includegraphics[width = \linewidth]{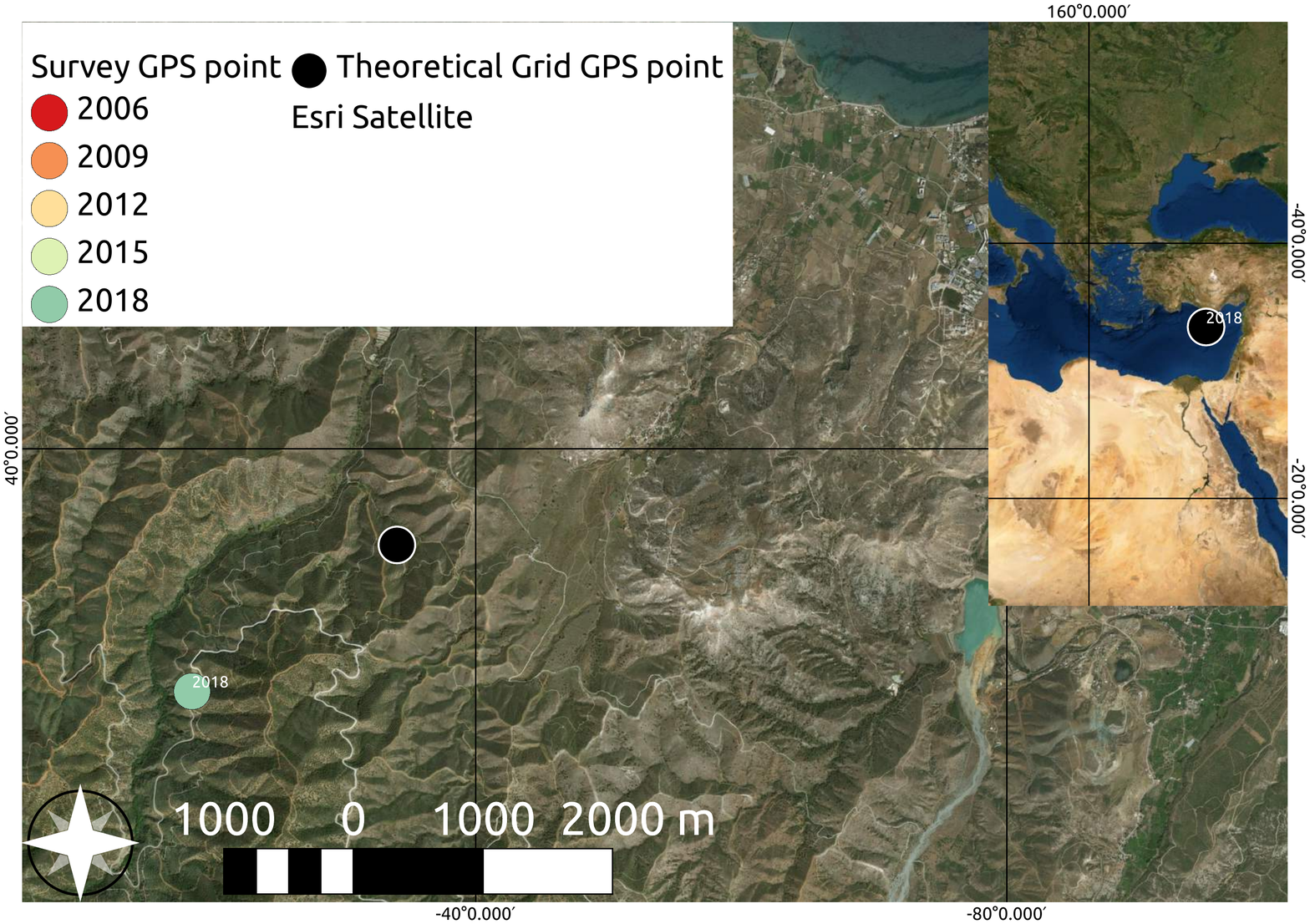}
    \caption{Example of a surveyed point (id 63861648) at large distance (1938m)}
  \end{subfigure}
\caption{The theoretical grid of LUCAS point location is stable in time, however, the observed location may vary according to the survey conditions, the terrain, and accuracy of the GPS positioning. }
\label{fig:LUCAS_map_distce}  
\end{figure}

The theoretical grid of LUCAS point location is stable in time, however, according to the survey conditions and the terrain and accuracy of the GPS positioning, the surveyor may not be able to reach the point. This results in effective variations of the position of the observer though time (Figure \ref{fig:LUCAS_map_distce} ).

%\begin{figure}[!!h]
%    \centering
%    \includegraphics[width=0.5\textwidth]{./figures/LUCAS_histo_th_gps_dist.png}
%    \caption{Distribution and metrics thereof of calculated distances between LUCAS THPs and GPSPs}
%    \label{fig:LUCAS_histo_th_gps_dist}
%\end{figure}

%\begin{figure}[!!h]
%    \centering
%    \includegraphics[width=\textwidth]{./figures/lucasConsistency.png}
%    \caption{Data quality checks and processing workflow}
%    \label{fig:lucasConst}
%\end{figure}

\section{Usage Notes}

%The Usage Notes should contain brief instructions to assist other researchers with reuse of the data. This may include discussion of software packages that are suitable for analysing the assay data files, suggested downstream processing steps (e.g. normalization, etc.), or tips for integrating or comparing the data records with other datasets. Authors are encouraged to provide code, programs or data-processing workflows if they may help others understand or use the data. Please see our code availability policy for advice on supplying custom code alongside Data Descriptor manuscripts.

%For studies involving privacy or safety controls on public access to the data, this section should describe in detail these controls, including how authors can apply to access the data, what criteria will be used to determine who may access the data, and any limitations on data use. 

%intro
To summarize, the work documented in this data descriptor consists of: 1) Multi-year harmonised LUCAS database, 2) Archive with high resolution LUCAS photos, 3) LUCAS survey geometries and point locations, 4) R package, 5) Data descriptor of resulting database and 6) a Documentation table for users to quickly identify the differences of collected data between LUCAS campaigns micro-data and harmonised database.

The harmonized LUCAS product greatly reduces the complexity and layered nature of the original LUCAS datasets. In doing so, it valorizes the effort of many surveyors, data cleaners, statisticians, and database maintainers.
%, and European taxpayers. 
The database's novelty lies in the fact that for the first time, users can query the whole LUCAS archive concurrently, allowing for comparisons and combinations between all variables collected for the relevant years. This homogeneity of the product facilitates the unearthing of temporal and spatial relations that were otherwise obscured by the physical separation between survey results. Moreover, by avoiding the burden of combing through the cumbersome documentation, the user is now free to concentrate on the research, thereby facilitating scientific discovery and analysis. The user-friendly nature of the database should make LUCAS data more usable as it enables easier access to ready-to-use information. This should enhance LUCAS data use and uptake by research institutes that support policy and decision making, educational organizations, NGOs, and (civil) society. 

%shortcomings
Naturally, the product suffers from the shortcomings inherent in the source data, such as possibly inadequate surveying, surveyor or technology-related errors of precision while taking coordinates or measurements, etc. The harmonisation process itself also reveals some inconsistencies in the source data. For instance, certain variables could not be harmonised between years. These are mostly related to measurements of percentage or extent of coverage, where in the early stages of LUCAS surveyors were asked to fill in a multiple choice question, listing a range of values, whereas in subsequent surveys the surveyor was asked to fill in the actual value in quantified units. This problem exists mostly, though not exclusively, for the 2006 survey. This makes it impossible for these variables to be translated into the user friendly version and these must thus remain in their original coding. For ease of identification we have marked them with a red asterisk (\textcolor{red}{*}) in the record descriptor (Table \ref{tab:descriptor}). Additional information can be found in the comments section of the record descriptor. 

Another shortcoming is the change of classification legends between the different surveys, mainly concerning LC/LU, as well as LC species and LU types. To document this shortcoming, a table is provided, see  special remarks in the Table (“LC (LU) changes” in the file \textit{LUCAS-Variable\_and\_Classification\_Changes.xlsx}, \url{https://figshare.com/s/4a2e5d119ee0a98bec6e}).

%what it can be used for + examples 

%pitfalls concerning using the data set for the use cases

\subsection{Code availability}
\label{sec:code}

%For all studies using custom code in the generation or processing of datasets, a statement must be included in the Methods section, under the subheading "Code availability", indicating whether and how the code can be accessed, including any restrictions to access. This section should also include information on the versions of any software used, if relevant, and any specific variables or parameters used to generate, test, or process the current dataset. 

To guarantee transparency and reproducibility, the harmonisation workflow was carried out with open-source tools, namely PostgreSQL (9.5.17)/PostGIS (2.1.8 r13775)) and R  (3.4.3) \cite{team2013r}). The code is provided as a R package containing 17 functions (downloadable here for the review \url{https://www.dropbox.com/sh/dwhoj1p4izds9bh/AABckSM_zNLZxzy613Yc5lIFa?dl=0} %and will be later pushed on CRAN
). The lucas package includes all the scripts and documentation (also provided in pdf). Additionally, along with the package, a script (\textit{main.R}) builds the harmonised data based step by step. The workflow is schematically shown in Figure \ref{fig:workflow} with all the functions in blue. All the processing is done with SQL with only column reordering and consistency checks being done in R. The code is freely available under GPL (>= 3) license.

\section{Acknowledgements}

%The Acknowledgements should contain text acknowledging non-author contributors. Acknowledgements should be brief, and should not include thanks to anonymous referees and editors or effusive comments. Grant or contribution numbers may be acknowledged.
The authors would like to acknowledge the many surveyors, quality controllers, and support personal who have been carrying out the LUCAS survey. Additionally, we would like to thank past and current members of the Eurostat LUCAS team (M. Kasanko, M. Fritz, H. Ramos, P. Jaques, C. Wirtz, L. Martino, ...) who have been instrumental over the years in implementing LUCAS at various stages. The authors would also like to thank N. Elvekjaer for her precious comments on the manuscript.

\section{Author contributions}

%Each author’s contribution to the work should be described briefly, on a separate line, in the Author Contributions section. 

B. E., A. P., P. D. are responsible of the LUCAS data collection. M.Y., R.D., G. L., L. M.-S. and M. v.d.V. processed and analyzed the data.
H. I. R. provides a storage solution to distribute the photos.
C. J. reviewed the DB and made the documentation table.
R.D., M.Y., G. L., B. E, A. P., P.D., J. G.,  H. I. R.,  L. M.-S., M. v.d.V. wrote the paper, provided comments and suggestions on the manuscript.

\section{Competing interests}

%A competing interests statement is required for all papers accepted by and published in \emph{Scientific Data}. If there is no conflict of interest, a statement declaring this must still be included in the manuscript.

The authors declare that they have no known competing financial interests or personal relationships that could have appeared to influence the work reported in this paper.

\section{Figures and figures legends}

% \begin{figure}[H]
%     \centering
%     \includegraphics[width=\textwidth]{./figures/lucas_harmo_revisit.eps}
%     \caption{Point revisit.}
%     \label{fig:lucas_harmo_revisit}
% \end{figure}

%Figure should be referred to using a consistent numbering scheme through the entire Data Descriptor. For initial submissions, authors may choose to supply this document as a single PDF with embedded figures, but separate figure image files must be provided for revisions and accepted manuscripts. In most cases, a Data Descriptor should not contain more than three figures, but more may be allowed when needed. We discourage the inclusion of figures in the Supplementary Information \textendash{} all key figures should be included here in the main Figure section. 

%Figure legends begin with a brief title sentence for the whole figure and continue with a short description of what is shown in each panel, as well as explaining any symbols used. Legend must total no more than 350 words, and may contain literature references. 

\section{Tables}

% % TABLE CHANGING NAME BEFORE ESTAT LAST CHANGES
% \begin{table}[H]
% \centering
%   \caption{Table of renamed variables}
%   \label{tab:changeName}
% \begin{tabular}{lll}
%   \hline
% \textbf{Year added to} & \textbf{Old name} & \textbf{Harmonised name} \\ 
%   \hline
%   2006 & surv\_date & surveydate \\ 
%   2006 & x\_laea & th\_lat \\ 
%   2006 & y\_laea & th\_long \\ 
%   2009, 2012, 2015 & area\_size & parcel\_area\_ha \\ 
%   2009, 2012, 2015 & date & surveydate \\ 
%   2009, 2012, 2015 & lc1\_pct & lc1\_perc \\ 
%   2009, 2012, 2015 & lc2\_pct & lc2\_perc \\ 
%   2009, 2012, 2015 & lc1\_species & lc1\_spec \\ 
%   2009, 2012, 2015 & lc2\_species & lc2\_spec \\ 
%   2009, 2012, 2015 & land\_mngt & grazing \\ 
%   2009, 2012, 2015 & obs\_dir & obs\_direct \\ 
%   2009, 2012, 2015 & photo\_e & photo\_east \\ 
%   2009, 2012, 2015 & photo\_w & photo\_west \\ 
%   2009, 2012, 2015 & photo\_n & photo\_north \\ 
%   2009, 2012, 2015 & photo\_s & photo\_south \\ 
%   2009, 2012, 2015 & photo\_p & photo\_point \\ 
%   2009, 2012, 2015 & soil\_plough & lndmng\_plough \\ 
%   2009, 2012 & tree\_height\_srv & tree\_height\_survey \\ 
%   2012, 2015 & tree\_height\_mat & tree\_height\_maturity \\ 
%   2015 & lu1\_pct & lu1\_perc \\ 
%   2015 & lu2\_pct & lu2\_perc \\ 
%   2015 & protected\_area & special\_status \\ 
%   \hline
% \end{tabular}
% \end{table}

% latex table generated in R 3.4.4 by xtable 1.8-4 package
% Tue Mar 31 19:41:09 2020
\begin{table}[ht]
\centering
  \caption{Table of renamed variables}
  \label{tab:changeName}
\begin{tabular}{lll}
  \hline
year.added.to & old.name & new.name \\ 
  \hline
2006 & surv\_date & surveydate \\ 
  2006 & x\_laea & th\_lat \\ 
  2006 & y\_laea & th\_long \\ 
  2009, 2012, 2015 & area\_size & parcel\_area\_ha \\ 
  2009, 2012, 2015 & date & surveydate \\ 
  2009, 2012, 2015 & lc1\_pct & lc1\_perc \\ 
  2009, 2012, 2015 & lc2\_pct & lc2\_perc \\ 
  2009, 2012, 2015 & lc1\_species & lc1\_spec \\ 
  2009, 2012, 2015 & lc2\_species & lc2\_spec \\ 
  2009, 2012, 2015 & land\_mngt & grazing \\ 
  2009, 2012, 2015 & obs\_dir & obs\_direct \\ 
  2009, 2012, 2015 & photo\_e & photo\_east \\ 
  2009, 2012, 2015 & photo\_w & photo\_west \\ 
  2009, 2012, 2015 & photo\_n & photo\_north \\ 
  2009, 2012, 2015 & photo\_s & photo\_south \\ 
  2009, 2012, 2015 & photo\_p & photo\_point \\ 
  2009, 2012, 2015 & tree\_height\_surv & tree\_height\_survey \\ 
 % 2009, 2012, 2015 & soil\_plough & lndmng\_plough \\ 
  2009, 2012, 2015 & soil\_stones & soil\_stones\_perc \\
  2012, 2015 & tree\_height\_mat & tree\_height\_maturity \\ 
  2015 & lu1\_pct & lu1\_perc \\ 
  2015 & lu2\_pct & lu2\_perc \\ 
  2015 & protected\_area & special\_status \\ 
  2015 & pi\_extension & office\_pi \\ 
   \hline
\end{tabular}
\end{table}

\begin{longtable}[H]{ll}

\caption{Table of added columns} 
\label{tab:addCols}
\footnotesize
\\
\hline
\textbf{Years added to} & \textbf{Variable} \\ 
\hline
\endfirsthead

\multicolumn{2}{c}%
{\tablename\ \thetable\ -- \textit{Continued from previous page}} \\
\hline
\textbf{Years added to} & \textbf{Variable} \\ 
\hline
\endhead

\hline \multicolumn{2}{c}{\textit{Continued on next page}} \\
\endfoot

\hline
\endlastfoot

  2006 & gps\_altitude \\ 
  2006, 2009, 2012 & lu1\_type \\ 
  2006, 2009, 2012 & lu1\_perc \\ 
  2006, 2009, 2012 & lu2\_type \\ 
  2006, 2009, 2012 & lu2\_perc \\ 
  2006, 2009, 2012 & inspire\_plcc1 \\ 
  2006, 2009, 2012 & inspire\_plcc2 \\ 
  2006, 2009, 2012 & inspire\_plcc3 \\ 
  2006, 2009, 2012 & inspire\_plcc4 \\ 
  2006, 2009, 2012 & inspire\_plcc5 \\ 
  2006, 2009, 2012 & inspire\_plcc6 \\ 
  2006, 2009, 2012 & inspire\_plcc7 \\ 
  2006, 2009, 2012 & inspire\_plcc8 \\ 
  2006, 2009, 2012, 2015 & nuts3 \\ 
  2006, 2009, 2012 & office\_pi \\ 
  2006, 2009, 2012, 2015 & ex\_ante \\ 
  2006, 2009, 2012, 2015 & car\_latitude \\ 
  2006, 2009, 2012, 2015 & car\_ew \\ 
  2006, 2009, 2012, 2015 & car\_longitude \\ 
  2009, 2012, 2015 & gps\_ew \\ 
  2006, 2009 & tree\_height\_maturity \\ 
  %2006, 2009, 2012, 2015 & lm\_plough\_slope \\ 
 % 2006, 2009, 2012, 2015 & lm\_plough\_direct \\ 
  2006, 2009, 2012, 2015 & lm\_stone\_walls \\ 
  2006, 2009, 2012, 2015 & lm\_grass\_margins \\ 
  2006, 2009 & special\_status \\ 
  2006, 2009 & lc\_lu\_special\_remark \\ 
  2006, 2009, 2012, 2015 & cprn\_cando \\ 
  2006, 2009, 2012, 2015 & cprn\_lc \\ 
  2006, 2009, 2012, 2015 & cprn\_lc1n \\ 
  2006, 2009, 2012, 2015 & cprnc\_lc1e \\ 
  2006, 2009, 2012, 2015 & cprnc\_lc1s \\ 
  2006, 2009, 2012, 2015 & cprnc\_lc1w \\ 
  2006, 2009, 2012, 2015 & cprn\_lc1n\_brdth \\ 
  2006, 2009, 2012, 2015 & cprn\_lc1e\_brdth \\ 
  2006, 2009, 2012, 2015 & cprn\_lc1s\_brdth \\ 
  2006, 2009, 2012, 2015 & cprn\_lc1w\_brdth \\ 
  2006, 2009, 2012, 2015 & cprn\_lc1n\_next \\ 
  2006, 2009, 2012, 2015 & cprn\_lc1e\_next \\ 
  2006, 2009, 2012, 2015 & cprn\_lc1s\_next \\ 
  2006, 2009, 2012, 2015 & cprn\_lc1w\_next \\ 
  2006, 2009, 2012, 2015 & cprn\_urban \\ 
  2006, 2009, 2012, 2015 & cprn\_impervious\_perc \\ 
  2006, 2009, 2012, 2015 & eunis\_complex \\ 
  2006, 2009, 2012, 2015 & grassland\_sample \\ 
  2006, 2009, 2012, 2015 & grass\_cando \\ 
  2006, 2009, 2012, 2015 & erosion\_cando \\ 
  2006, 2009, 2012, 2015 & bio\_sample \\ 
  2006, 2009, 2012, 2015 & soil\_bio\_taken \\ 
  2006, 2009, 2012, 2015 & bulk0\_10\_sample \\ 
  2006, 2009, 2012, 2015 & soil\_blk\_0\_10\_taken \\ 
  2006, 2009, 2012, 2015 & bulk10\_20\_sample \\ 
  2006, 2009, 2012, 2015 & soil\_blk\_10\_20\_taken \\ 
  2006, 2009, 2012, 2015 & bulk20\_30\_sample \\ 
  2006, 2009, 2012, 2015 & soil\_blk\_20\_30\_taken \\ 
  2006, 2009, 2012, 2015 & standard\_sample \\ 
  2006, 2009, 2012, 2015 & soil\_std\_taken \\ 
  2006, 2009, 2012, 2015 & organic\_sample \\ 
  2006, 2009, 2012, 2015 & soil\_org\_depth\_cando \\ 
  2006, 2009, 2012, 2015 & crop\_residues \\ 
  2018 & obs\_radius \\ 
  2018 & soil\_taken \\ 
  2018 & soil\_crop \\ 
  2006, 2018 & transect \\ 
  ALL & letter\_group \\ 
  ALL & file\_path\_thumb\_n \\ 
  ALL & file\_path\_thumb\_s \\ 
  ALL & file\_path\_thumb\_w \\ 
  ALL & file\_path\_thumb\_e \\ 
  ALL & file\_path\_thumb\_p \\ 
  ALL & year \\ 
\hline
\end{longtable}

% TABLE OF NEWLY ADDED COLUMNS
% latex table generated in R 3.6.2 by xtable 1.8-4 package
% Thu Jan 16 14:23:44 2020
\begin{table}[H]
\centering
\caption{Table of newly added columns}
\label{tab:new_cols}
\begin{tabular}{rll}
  \hline
\textbf{Column name}  & \textbf{Description}  \\ 
  \hline
letter\_group & First level of LUCAS LC1/2 classification \\ 
year & Year of the survey \\ 
file\_path\_gisco\_n/s/e/w/p & Path to cardinal or point photo on GISCO cloud \\ 
th\_geom & Geometry of theoretical LUCAS point according to grid \\ 
gps\_geom & Geomtery at the point the surveyor reached \\ 
th\_gps\_dist & Calculated distance between the two points \\ 
revisit & Numbers of years of revisit for the LUCAS point \\ 
   \hline
\end{tabular}
\end{table}

\newpage
\begin{longtable}[H]{@{}p{3cm}p{0.5cm}p{0.5cm}p{2.2cm}p{0.7cm}p{0.7cm}p{2.5cm}@{}}
   \caption{Table of re-coded variables}
   \label{tab:recoding}
\footnotesize
\\
\hline
 \textbf{Variable} & \textbf{Year} &\textbf{Old coding} & \textbf{Value} & \textbf{Interim} & \hspace{0.5cm} \textbf{Harmo} & \textbf{Value}\\
\hline
\endfirsthead

\multicolumn{7}{c}%
{\tablename\ \thetable\ -- \textit{Continued from previous page}} \\
\hline
 \textbf{Variable} & \textbf{Year} &\textbf{Old coding} & \textbf{Value} & \textbf{Interim} & \hspace{0.5cm} \textbf{Harmo} & \textbf{Value} \\  
\hline
\endhead

\hline \multicolumn{4}{c}{\textit{Continued on next page}} \\
\endfoot

\hline
\endlastfoot
  \hline
  soil\_taken & 2009 &   3 & Point not in soil sample & - &   4 & No sample required \\ 
   &  &  &  &  &  &  \\ 
  lc1/2\_perc & 2012 &   2 & 5 – 10;  & - &   1 & $<$ 10;  \\ 
  lc1/2\_perc & 2012 &   3 & 10 – 25;  & - &   2 & 10 – 25;  \\ 
  lc1/2\_perc & 2012 &   4 & 25 – 50;  & - &   3 & 25 – 50;  \\ 
  lc1/2\_perc & 2012 &   5 & 50 – 75; & - &   4 & 50 – 75;  \\ 
  lc1/2\_perc & 2012 &   6 & $>$ 75; & - &   5 & $>$ 75; \\ 
   &  &  &  &  &  &  \\ 
  lc\_lu\_
  
  special\_remark & 2012 &   1 & Tilled/Sowed & 100 &   2 & Tilled/sowed \\ 
  lc\_lu\_
  
  special\_remark & 2012 &   2 & Harvested field & 200 &   1 & Harvested field \\ 
  lc\_lu\_
  
  special\_remark & 2012 &   7 & No remark & 700 &  10 & No remark \\ 
  lc\_lu\_
  
  special\_remark & 2012 &   8 & Not relevant & 800 &  88 & Not relevant \\ 
   &  &  &  &  &  &  \\ 
  obs\_type & 2015 &   5 & Marine sea & - &   8 & Marine sea \\ 
  obs\_type & 2015 &   6 & Out of national territory & - &   5 & Out of national territory \\ 
   &  &  &  &  &  &  \\ 
  lc\_lu\_
  
  special\_remark & 2015 &   1 & Tilled/Sowed & 100 &   2 & Tilled/Showed \\ 
  lc\_lu\_
  
  special\_remark & 2015 &   2 & Harvested field & 200 &   1 & Harvested field \\ 
  lc\_lu\_
  
  special\_remark & 2015 &   7 & No remark & 700 &  10 & No remark \\ 
  lc\_lu\_
  
  special\_remark & 2015 &   8 & Not relevant & 800 &  88 & Not relevant \\ 
  lc\_lu\_
  
  special\_remark & 2015 &   9 & Temporarily dry & 900 &   8 & Temporarily dry \\ 
  lc\_lu\_
  
  special\_remark & 2015 &  10 & Temp flooded & 1000 &   9 & Temp flooded \\ 
   &  &  &  &  &  &  \\ 
  lc1/2\_perc & 2015 &   2 & 5 – 10;  & - &   1 & $<$ 10;  \\ 
  lc1/2\_perc & 2015 &   3 &  10 – 25;  & - &   2 & 10 – 25;  \\ 
  lc1/2\_perc & 2015 &   4 & 25 – 50;  & - &   3 & 25 – 50;  \\ 
  lc1/2\_perc & 2015 &   5 & 50 – 75; & - &   4 & 50 – 75;  \\ 
  lc1/2\_perc & 2015 &   6 & 75 – 90; & - &   5 & $>$ 75; \\ 
  lc1/2\_perc & 2015 &   7 & $>$ 90; & - &   5 & $>$ 75; \\ 
   &  &  &  &  &  &  \\ 
  osb\_type & 2018 &   6 & Out of EU28 & - &   5 & Out of national territory \\ 
   &  &  &  &  &  &  \\ 
  parcel\_area\_ha & 2018 &   2 & 0.1 – 0.5; & - &   1 & $<$ 0.5; \\ 
  parcel\_area\_ha & 2018 &   3 & 0.5 – 1; & - &   2 &  0.5 – 1; \\ 
  parcel\_area\_ha & 2018 &   4 & 1 – 10; & - &   3 &  1 – 10;  \\ 
  parcel\_area\_ha & 2018 &   5 & $>$ 10; & - &   4 & $>$ 10; \\ 
   \hline

\end{longtable}

\begin{longtable}[H]{p{0.02\textwidth}p{0.18\textwidth}p{0.4\textwidth}p{0.4\textwidth}p{0.01\textwidth}}
\caption{Data descriptor of the resulting database} 
\label{tab:descriptor}
\footnotesize
\\
\hline
 & \textbf{Variable} & \textbf{Values} & \textbf{Description} & \textbf{Collection year} \\ 
\hline
\endfirsthead

\multicolumn{5}{c}%
{\tablename\ \thetable\ -- \textit{Continued from previous page}} \\
\hline
 & \textbf{Variable} & \textbf{Values} & \textbf{Description} & \textbf{Collection year} \\ 
\hline
\endhead

\hline \multicolumn{5}{c}{\textit{Continued on next page}} \\
\endfoot

\hline
\endlastfoot
1 & id & - & Table identifier & all \\ 
  2 & point\_id & - & Unique point identifier & all \\ 
  3 & year & - & Which year the point was surveyed & all \\ 
  4 & nuts0 & - & NUTS 2016 Level 0 & all \\ 
  5 & nuts1 & - & NUTS 2016 Level 1 & all \\ 
  6 & nuts2 & - & NUTS 2016 Level 2 & all \\ 
  7 & nuts3 & - & NUTS 2016 Level 3 & 2018 \\ 
  8 & th\_lat & - & Theoretical latitude (WGS84) of the LUCAS point according to the LUCAS grid & all \\ 
  9 & th\_long & - & Theoretical longitude (WGS84) of the LUCAS point according to the LUCAS grid & all \\ 
  10 & office\_pi & 0 - No; 
1 - Yes; & Indication of whether photo-interpretation has happened in the office for this LUCAS point & 2015 2018 \\ 
  11 & ex\_ante & 0 - No; 
1 - Yes; &  & 2018 \\ 
  12 & survey\_date & - & Date on which the survey was carried out & all \\ 
  13 & car\_latitude & - & Latitude (WGS84) on which the car was parked & 2018 \\ 
  14 & car\_ew & 1 - East; 
2 - West; 
8 - Not relevant; & GPS Car parking East/West & 2018 \\ 
  15 & car\_longitude & - & Longitude (WGS84) on which the car was parked & 2018 \\ 
  16 & gps\_proj & 1 - WGS84; 
2 - No GPS signal; 
8 - Not relevant; & Normal functioning of GPS using “WGS 84” as coordinate system. & all \\ 
  17 & gps\_prec & - & Indication of average location error as given by GPS receiver (in meters) & all \\ 
  18 & gps\_altitude & - & Elevation in m above sea level & 2009 2012 2015 2018 \\ 
  19 & gps\_lat & - & GPS latitude of the location from which observation is actually done (WGS84) & all \\ 
  20 & gps\_ew & 1 - East; 
2 – West;
8 – Not relevant; & East-west encoding setting for GPS. & all \\ 
  21 & gps\_long & - & GPS longitude of the location from which observation is actually done (WGS84) & all \\ 
  22 & obs\_dist & - & Indication of the distance between observation location and the LUCAS point as provided by the GPS receiver (in meters). & all \\ 
  23 & obs\_direct & 1 - On the point; 
2 - Look to the North; 
3 - Look to the East; 
8 - Not relevant; & 1 - On the point Point regularly observed.
2 – To North “Look to the North” rule is applied if the point is located on a boundary/ edge or a small linear feature ($<$3 m wide).
3 – To East "Look to the East" rule is applied if the point is located on a boundary/edge or a small linear feature ($<$3 m wide) directed.
8 – Not relevant & all \\ 
  24 & obs\_type & 1 - In situ $<$ 100 m; 
2 - In situ $>$ 100 m; 
3 - In situ PI; 
4 - In situ PI not possible; 
5 - Out of national territory; 
7 - In office PI;
8 – Marine Sea; (only 2015) & The method of observation for the relevant point.  & all \\ 
  25 & obs\_radius & 1 – 1.5;
2 - 20; 
8 - Not relevant; & The radius of observation – whether the immediate or the extended window of observation is taken under consideration. & 2006 2009 2012 2015 \\ 
  26 & letter\_group & - & Which letter group (top tier classification) from C3 does the point belong to & all \\ 
  27 & lc1 & - & Coding of land cover according to LUCAS 2018 classification. & all \\ 
  28 & lc1\_label & - & Label of field lc1 & all \\ 
  29 & lc1\_spec & - & Coding of land cover species according to LUCAS 2018 classification. & 2009 2012 2015 2018 \\ 
  30 & lc1\_spec\_label & - & Label of field lc1\_spec & all \\ 
  31 & lc1\_perc \textcolor{red}{*} & 1 - $<$ 10; 
2 - 10 - 25; 
3 - 25 - 50; 
4 - 50 - 75; 
5 - $>$ 75;
8 – Not relevant; & The percentage that the land cover (lc1) takes on the ground.  & 2009 2012 2015 2018 \\ 
  32 & lc2 & - & Coding of land cover according to LUCAS 2018 classification. & all \\ 
  33 & lc2\_label & - & Label of field lc2 & all \\ 
  34 & lc2\_spec & - & Coding of land cover species according to LUCAS 2018 classification. & 2009 2012 2015 2018 \\ 
  35 & lc2\_spec\_label & - & Label of field lc2\_spec & all \\ 
  36 & lc2\_perc \textcolor{red}{*} & 1 - $<$ 10; 
2 - 10 - 25; 
3 - 25 - 50; 
4 - 50 - 75; 
5 - $>$ 75;
8 – Not relevant; & The percentage that the land cover (lc2) takes on the ground. & 2009 2012 2015 2018 \\ 
  37 & lu1 & - & Coding of the land use according to LUCAS LU 2018 classification & all \\ 
  38 & lu1\_label & - & Label of field lu1 & all \\ 
  39 & lu1\_type & - & Coding of the land use types according to LUCAS LU 2018 classification & 2015 2018 \\ 
  40 & lu1\_type\_label & - & Label of field lu1\_type & all \\ 
  41 & lu1\_perc \textcolor{red}{*} & 1 -  $<$ 5;
2 - 5 – 10;
3 - 10 – 25;
4 - 25 – 50;
5 - 50 – 75;
6 - 75 – 90;
7 -  $\geq$ 90;
8 – Not Relevant; & The percentage that the land use (lu1) takes on the ground. & 2015 2018 \\ 
  42 & lu2 & - & Coding of the land use according to LUCAS LU 2018 classification & all \\ 
  43 & lu2\_label & - & Label of field lu2 &  \\ 
  44 & lu2\_type & - & Coding of the land use types according to LUCAS LU 2018 classification & 2015 2018 \\ 
  45 & lu2\_type\_label & - & Label of field lu2\_type & all \\ 
  46 & lu2\_perc \textcolor{red}{*} & 1 -  $<$ 5;
2 - 5 – 10;
3 - 10 – 25;
4 - 25 – 50;
5 - 50 – 75;
6 - 75 – 90;
7 - $\geq$ 90;
8 - Not Relevant; & The percentage that the land use (lu2) takes on the ground.  & 2015 2018 \\ 
  47 & parcel\_area\_ha & 1 - $<$ 0.5;
2 - 0.5 – 1;
3 - 1 – 10;
4 - $>$ 10;
8 – Not Relevant; & Size of the surveyed parcel in hectares. & 2009 2012 2015 2018 \\ 
  48 & tree\_height\_
  
  maturity & 1 - $<$ 5 m ; 
2 - $>$ 5 m; 
8 - Not relevant; 
255 – Not identifiable; & Height of trees at maturity & 2012 2015 2018 \\ 
  49 & tree\_height\_
  
  survey & 1 - $<$ 5 m ; 
2 - $>$ 5 m; 
8 - Not relevant; 
255 – Not identifiable; & Height of trees at the moment of survey & 2009 2012 2015 2018 \\ 
  50 & feature\_width & 1 - $<$ 20 m; 
2 - $>$ 20 m; 
8 - Not relevant;
255 – Not identifiable; & Width of the feature & 2009 2012 2015 2018 \\ 
  51 & lm\_stone\_walls & 1 - No; 
2 - Stone wall not maintained; 
3 - Stone wall well maintained; 
8 - Not relevant; & Presence of stone walls on the plot.  & 2018 \\ 
  52 & crop\_residues & 1 - Yes; 
2 - No;
8 - Not relevant; & Presence of crop residues on the plot & 2018 \\ 
  53 & lm\_grass\_
  
  margins & 1 - No; 
2 - $<$ 1 m width; 
3 - $>$ 1 m width; 
8 - Not relevant; & Presence of grass margins on the plot.  & 2018 \\ 
  54 & grazing & 1 - Signs of grazing; 
2 - No signs of grazing; 
8 - Not relevant;  & Signs of grazing on the plot. & 2009 2012 2015 2018 \\ 
  55 & special\_status & 1 - Protected; 
2 - Hunting; 
3 - Protected and hunting; 
4 - No special status; 
8 - Not relevant; & Whether the plot is part of any specially regulated area.  & 2012 2015 2018 \\ 
  56 & lc\_lu\_
  
  special\_remark & 1 - Harvested field; 
2 - Tilled/sowed; 
3 - Clear cut; 
4 - Burnt area; 
5 - Fire break; 
6 - Nursery; 
7 - Dump site; 
8 - Temporary dry; 
9 - Temporary flooded; 
10 - No remark; 
88 - Not Relevant; & Any special remarks on the land cover / land use. & 2012 2015 2018 \\ 
  57 & cprn\_cando & 1 - Yes; 
2 - No; 
8 - Not relevant; & Can you do a Copernicus survey on this point? & 2018 \\ 
  58 & cprn\_lc & - & The land cover on the Copernicus points according to the classification scheme at level2 & 2018 \\ 
  59 & cprn\_lc\_label & - & Label of field cprn\_lc & all \\ 
  60 & cprn\_lc1n & - & The extent (in meters) to which the land cover of the Copernicus point stays the same in direction North & 2018 \\ 
  61 & cprnc\_lc1e & - & The extent (in meters) to which the land cover of the Copernicus point stays the same in direction East & 2018 \\ 
  62 & cprnc\_lc1s & - & The extent (in meters) to which the land cover of the Copernicus point stays the same in direction South & 2018 \\ 
  63 & cprnc\_lc1w & - & The extent (in meters) to which the land cover of the Copernicus point stays the same in direction West & 2018 \\ 
  64 & cprn\_lc1n\_
  
  brdth & - & The breath (in \%) to the next Copernicus land cover in direction North & 2018 \\ 
  65 & cprn\_lc1e\_
  
  brdth & - & The breath (in \%) to the next Copernicus land cover in direction East & 2018 \\ 
  66 & cprn\_lc1s\_
  
  brdth & - & The breath (in \%) to the next Copernicus land cover in direction South & 2018 \\ 
  67 & cprn\_lc1w\_
  
  brdth & - & The breath (in \%) to the next Copernicus land cover in direction West & 2018 \\ 
  68 & cprn\_lc1n\_next & - & The next Copernicus land cover (level2) in direction North & 2018 \\ 
  69 & cprn\_lc1s\_next & - & The next Copernicus land cover (level2) in direction South & 2018 \\ 
  70 & cprn\_lc1e\_next & - & The next Copernicus land cover (level2) in direction East & 2018 \\ 
  71 & cprn\_lc1w\_next & - & The next Copernicus land cover (level2) in direction West & 2018 \\ 
  72 & cprn\_urban & 1 - Yes; 
2 - No; 
8 - Not relevant & Is the Copernicus point located in an urban area. & 2018 \\ 
  73 & cprn\_impervious\_
  
  perc & - & Assess the percentage of impervious surfaces & 2018 \\ 
  74 & inspire\_plcc1 & - & Assess the percentage of coniferous trees & 2015 2018 \\ 
  75 & inspire\_plcc2 & - & Assess the percentage of broadleaved trees & 2015 2018 \\ 
  76 & inspire\_plcc3 & - & Assess the percentage of shrubs & 2015 2018 \\ 
  77 & inspire\_plcc4 & - & Assess the percentage of herbaceous plants & 2015 2018 \\ 
  78 & inspire\_plcc5 & - & Assess the percentage of lichens and mosses & 2015 2018 \\ 
  79 & inspire\_plcc6 & - & Assess the percentage of consolidated (bare) surface (e.g. rock outcrops) & 2015 2018 \\ 
  80 & inspire\_plcc7 & - & Assess the percentage of unconsolidated (bare) surface (e.g. sand) & 2015 2018 \\ 
  81 & inspire\_plcc8 & - & Sum of all classes must be 100\%. This field covers for the difference, if it exists. & 2015 2018 \\ 
  82 & eunis\_complex & 6 - X06; 
9 - X09; 
10 - Other; 
11 - Unknown; 
88 - Not relevant & EUNIS habitat classification & 2018 \\ 
  83 & grassland\_
  
  sample & 0 - TRUE; 
1 - False; & Whether or not the point is part of the grassland module & 2018 \\ 
  84 & grass\_cando & 1 - Yes; 
2 - No;
8 - Not relevant; & Is a grassland survey possible? & 2018 \\ 
  85 & wm & 1 - Irrigation; 
2 - Potential irrigation; 
3 - Drainage; 
4 - Irrigation and drainage; 
5 - No visible water management; 
8 - Not relevant;  & What type of water management is present at the point & 2009 2012 2015 2018 \\ 
  86 & wm\_source & 1 - Well; 
2 - Pond/Lake/Reservoir; 
3 - Stream/Canal/Ditch; 
4 - Lagoon/Wastewater; 
5 - Other/Not Identifiable; 
6 - Combo - Pond/Lake/Reservoir + Stream/Canal/Ditch; 
8 - Not relevant; 
16 - Other/Not Identifiable; 
17 - Combo - Other/Not Identifiable + Well;
18 - Combo - Other/Not Identifiable + Pond/Lake/Reservoir; 
20 - Combo - Other/Not Identifiable + Stream/Canal/Ditch; 
24 - Combo - Other/Not Identifiable + Lagoon/Wastewater; & What is the source of the irrigation at the point & 2009 2012 2015 2018 \\ 
  87 & wm\_type & 1 - Gravity; 
2 - Pressure: Sprinkle irrigation; 
3 - Pressure: Micro-irrigation; 
4 - Gravity/Pressure; 
5 - Other/Non identifiable; 
6 - Combo - Pressure: Sprinkle irrigation + Pressure: Micro-irrigation; 
8 - Not relevant; 
9 - Combo - Gravity/Pressure + Gravity; 
10 - Combo - Pressure: Sprinkle irrigation + Gravity/Pressure; 
12 - Combo - Pressure: Micro-irrigation + Gravity/Pressure; 
16 - Other/not identifiable; 
17 - Combo - Other/Non identifiable + Gravity; 
18 - Combo - Other/Non identifiable + Gravity/Pressure; 
24 - Combo - Other/Non identifiable + Gravity/Pressure + Pressure: Micro-irrigation; & The type of irrigation present at the point & 2009 2012 2015 2018 \\ 
  88 & wm\_delivery & 1 - Canal; 
2 - Ditch; 
3 - Pipeline; 
4 - Other/Non identifiable; 
5 - Other/Non identifiable + Canal; 
6 - Combo - Pipeline + Ditch; 
8 - Not relevant; 
10 - Combo - Other/Non identifiable + Ditch; 
12 - Combo - Other/Non identifiable + Pipeline; & The irrigation delivery system at the point & 2009 2012 2015 2018 \\ 
  89 & erosion\_cando & 1 - Yes; 
2 - No; 
8 - Not relevant; & Indicates whether a point is to be considered for assessing erosion (Yes) or not (No) & 2018 \\ 
  90 & soil\_stones\_
  
  perc \textcolor{red}{*} & 1 - $<$ 10; 
2 - 10 - 25; 
3 - 25 - 50; 
4 - $>$ 50;
8 – Not relevant; & Indicate the percentage of stones on the surface (does not include stones covered by soil) & 2009 2012 2015 2018 \\ 
  91 & bio\_sample & 0 - True; 
1 - False & Is the point a biodiversity sample point? & 2018 \\ 
  92 & soil\_bio\_taken & 0 - True; 
1 - False;
8 - Not relevant; & Was a soil-biodiversity sample taken? & 2018 \\ 
  93 & bulk0\_10\_
  
  sample & 0 - True; 
1 - False & Indicates whether a point is to be considered for collecting the bulk density between the given range & 2018 \\ 
  94 & soil\_blk\_
  
  0\_10\_taken & 1 - Yes; 
2 – No;
8 - Not relevant; & Has the soil sample between the given range been taken? & 2018 \\ 
  95 & bulk10\_
  
  20\_sample & 0 - True; 
1 - False & Indicates whether a point is to be considered for collecting the bulk density between the given range & 2018 \\ 
  96 & soil\_blk\_
  
  10\_20\_taken & 1 - Yes; 
2 – No;
8 - Not relevant; & Has the soil sample between the given range been taken? & 2018 \\ 
  97 & bulk20\_
  
  30\_sample & 0 - True; 
1 - False & Indicates whether a point is to be considered for collecting the bulk density between the given range & 2018 \\ 
  98 & soil\_blk\_
  
  20\_30\_taken & 1 - Yes; 
2 – No;
8 - Not relevant; & Has the soil sample between the given range been taken? & 2018 \\ 
  99 & standard\_
  
  sample & 0 - True; 
1 - False & Is the point is a standard soil point? & 2018 \\ 
  100 & soil\_std\_taken & 1 - Yes; 
2 – No;
8 - Not relevant; & Is the standard soil sample was taken? & 2018 \\ 
  101 & organic\_sample & 0 - True; 
1 - False & Is the point the point an organic sample point? & 2018 \\ 
  102 & soil\_org\_
  
  depth\_cando & 1 - Yes; 
2 – No;
8 - Not relevant; & Can depth be evaluated? & 2018 \\ 
  103 & soil\_taken & 1 – Yes
2 – Not possible
3 – No, already taken
4 – No sample required
8 – Not Relevant & Has a soil sample been taken (before 2018) & 2009 2012 2015 \\ 
  104 & soil\_crop & 1 - $<$ 10; 
2 - 10 - 25; 
3 - 25 - 50; 
4 - $>$ 50; 
8 - Not relevant; & Percentage of residual crop (only 2015) & 2009 2012 2015 \\ 
  105 & photo\_point & 1 - Yes; 
2 – No;
8 - Not relevant; & Has a photo on the point been taken? & all \\ 
  106 & photo\_north & 1 - Yes; 
2 – No;
8 - Not relevant; & Has a photo looking north been taken? & all \\ 
  107 & photo\_south & 1 - Yes; 
2 – No;
8 - Not relevant; & Has a photo looking south been taken? & all \\ 
  108 & photo\_east & 1 - Yes; 
2 – No;
8 - Not relevant; & Has a photo looking east been taken? & all \\ 
  109 & photo\_west & 1 - Yes; 
2 – No;
8 - Not relevant; & Has a photo looking west been taken? & all \\ 
  110 & transect & - & The changes in landcover as recorded by the 250 m east-facing transect line & 2009 2012 2015 \\ 
  111 & revisit & - & Number of years for which the point has been revisited  & all \\ 
  112 & th\_gps\_dist & - & Calculated distance between the theoretical location of the LUCAS point according to the grid and the actual recorded GPS location & all \\ 
  113 & file\_path\_
  
  gisco\_north & - & File path to north-facing image as stored on ESTAT GISCO sever. & all \\ 
  114 & file\_path\_
  
  gisco\_south & - & File path to south-facing image as stored on ESTAT GISCO sever. & all \\ 
  115 & file\_path\_
  
  gisco\_east & - & File path to east-facing image as stored on ESTAT GISCO sever. & all \\ 
  116 & file\_path\_
  
  gisco\_west & - & File path to west-facing image as stored on ESTAT GISCO sever. & all \\ 
  117 & file\_path\_
  
  gisco\_point & - & File path to point-facing image as stored on ESTAT GISCO sever. & all \\ 
   \hline 

\end{longtable}

%Authors are encouraged to provide one or more tables that provide basic information on the main ‘inputs’ to the study (e.g. samples, participants, or information sources) and the main data outputs of the study; also see the additional information on providing metadata on page 6. Tables in the manuscript should generally not be used to present primary data (i.e. measurements). Tables containing primary data should be submitted to an appropriate data repository.

%Tables may be provided within the \LaTeX{} document or as separate files (tab-delimited text or Excel files). Legends, where needed, should be included here. Generally, a Data Descriptor should have fewer than ten Tables, but more may be allowed when needed. Tables may be of any size, but only Tables which fit onto a single printed page will be included in the PDF version of the article (up to a maximum of three). 

%Due to typesetting constraints, tables that do not fit onto a single A4 page cannot be included in the PDF version of the article and will be made available in the online version only. Any such tables must be labelled in the text as ‘Online-only’ tables and numbered separately from the main table list e.g. ‘Table 1, Table 2, Online-only Table 1’ etc.

% \externalbibliography{yes}
\bibliographystyle{unsrt}
\bibliography{bibliography.bib}

\subsection*{Citing Data}
%In line with emerging industry-wide standards for data citation, references to all datasets described or used in the manuscript should be cited in the text with a superscript number and listed in the ‘References’ section in the same manner as a conventional literature reference. See the examples above.

\end{document}